\documentclass[%
reprint,
superscriptaddress,
 amsmath,amssymb,
 aps,
prd,
]{revtex4-2}

\usepackage{graphicx}
\usepackage{dcolumn}
\usepackage{bm}

\begin{document}


\title{Charactering instrumental noises and stochastic gravitational wave signals from combined time-delay interferometry}


\author{Gang Wang}
\email[Gang Wang: ]{gwang@shao.ac.cn, gwanggw@gmail.com}
\affiliation{Shanghai Astronomical Observatory, Chinese Academy of Sciences, Shanghai, 200030, China}

\author{Bin Li}
\email[Bin Li: ]{binli@pmo.ac.cn}
\affiliation{Purple Mountain Observatory, Chinese Academy of Sciences, Nanjing, 210023, China}
\affiliation{University of Science and Technology of China, Hefei, 230026, China}

\author{Peng Xu}
\email[Peng Xu: ]{xupeng@imech.ac.cn}
\affiliation{Institute of Mechanics, Chinese Academy of Sciences, Beijing 100190, China}
\affiliation{Lanzhou Center for Theoretical Physics, Lanzhou University, Lanzhou 730000, China}
\affiliation{Hangzhou Institute for Advanced Study, University of Chinese Academy of Sciences, Hangzhou 310124, China}

\author{Xilong Fan} 
\email[Xilong Fan: ]{xilong.fan@whu.edu.cn}
\affiliation{School of Physics and Technology, Wuhan University, Wuhan, Hubei 430072, China}

\date{\today} 

\begin{abstract}

LISA will detect gravitational waves (GWs) in the milli-Hz frequency band in space. Time-delay interferometry (TDI) is developed to suppress laser frequency noise beneath the acceleration noise and optical metrology noise. To identify stochastic GW signals, it would be required to characterize these noise components entangled in TDI data streams. In this work, we investigate noises characterization by combining the first-generation TDI channels from Michelson and Relay configurations. The Michelson channels are helpful to characterize acceleration noises in the lower frequency band, and the Relay configuration could effectively resolve optical path noises in the higher frequencies. Synergy could be achieved from their combination to determine these instrumental noises. Based on the characterized noises, we further reconstruct the power spectrum of noise in the selected TDI channel. Two cases are performed to characterize the spectrum shape of a stochastic GW signal. For a modeled signal, its parameter(s) could be directly estimated from the TDI data, and its spectrum could be recovered from the inferred values. And for an unexpected signal, its spectrum may be recognized and retrieved from noise-subtracted residual in which its power spectral density surpasses the noise level.

\end{abstract}

\maketitle

\section{Introduction}

The Laser Interferometer Space Antenna (LISA) is planned to observe gravitational waves (GW) in the milli-Hz frequency band. Three spacecraft are employed to orbit the Sun and form triangular interferometers with an arm length of $2.5 \times 10^6$ km \cite{2017arXiv170200786A}. The drag-free technology is utilized to keep test masses on spacecraft following their geodesics, and laser metrology is implemented to measure distance variation between test masses yielded by GWs. Due to the perturbations of planets, the arm length of the interferometers could not be fully equal and vary with time. Also because of the long baseline and capability of the laser source, the laser frequency noise will overwhelm the GW signals if an original Michelson laser interferometer is implemented. Time delay interferometry (TDI) is developed for the LISA mission to suppress the laser noise and achieve targeting sensitivity \cite{1999ApJ...527..814A,2000PhRvD..62d2002E}. Two generations of TDI have been developed to overcome the laser noise in different conditions. The first-generation TDI is designed to suppress the laser noise in a static unequal-arm interferometer \cite[and references therein]{1999ApJ...527..814A,2000PhRvD..62d2002E,2001CQGra..18.4059A,Larson:2002xr,Dhurandhar:2002zcl,Prince:2002hp,2003PhRvD..67l2003T,Vallisneri:2004bn,2008PhRvD..77b3002P,Tinto:2020fcc}, and the second-generation is considered to further cancel the laser frequency noise in a time-varying triangular configuration up to the first-order derivative with respect to time \cite[and references therein]{Shaddock:2003dj,Cornish:2003tz,Tinto:2003vj,Vallisneri:2005ji,Dhurandhar:2010pd,Tinto:2018kij,2019PhRvD..99h4023B,2020arXiv200111221M,Vallisneri:2020otf,Wang:2ndTDI}.

The principle of TDI is to combine multiple time-shifted laser links and form an equivalent equal-arm interferometer. The laser frequency noise could be effectively suppressed in the (closely) equal interferometric paths. With the cancellation of laser noise, the acceleration noise and optical path noise become the dominant detection noises. Since three spacecraft are involved in TDI, its data will contain the instrumental noises from multiple optical benches and test masses. If the performances of six test masses and optical benches are fully identical, the acceleration noise and optical path noise could be precisely characterized by the first-generation Michelson observables \cite{Caprini:2019pxz,Flauger:2020qyi,Boileau:2020rpg}. However, in the realistic case, the performance of instruments on each spacecraft may differ from each other, and these noises could be characterized individually by using three optimal channels \cite{Adams:2010vc,Adams:2013qma}, null streams \cite{Muratore:2021uqj}, or the observables derived from basic Sagnac generators \cite{Hartwig:2021mzw}.

The accuracy of noise characterization could affect the GW identifications, especially for a stochastic signal. For the galactic foreground or a stochastic gravitational wave background (SGWB), the signal persists in the data and is difficult to be discriminated from noise in a single detector. The galactic foreground is generated from overlapped GWs emitted from numerous compact binaries in Milky Way. Since most of the binaries could not be resolved by LISA, the foreground would be a confusing noise and affect the sensitivity at $\sim$1 mHz \cite{Cornish:2017vip,Korol:2017qcx}. The SGWB could be an astrophysical origin yielded by the unresolved compact binary systems out of the galaxy, or be cosmological origin which generated by the mechanisms in the early Universe \cite{LIGOScientific:2019vic,Romano:2016dpx,Hindmarsh:2013xza,Caprini:2015zlo,Bartolo:2016ami,Caprini:2019egz,Caprini:2019pxz,Flauger:2020qyi,Christensen:2018iqi}. The searchings for SGWB from the advanced LIGO and advanced Virgo observations are actively ongoing \cite{TheLIGOScientific:2016dpb,Abbott:2021xxi,Abbott:2021ksc,Abbott:2021jel}. For the LISA mission, with knowing the spectral shapes of signals, the SGWB could be identified and reconstructed from targeting searches \cite{Caprini:2019pxz,Flauger:2020qyi,Boileau:2020rpg}. \citet{Tinto:2001ii} explored the instrumental noise and SGWB discrimination by considering the Michelson and symmetric Sagnac observables, and \citet{Adams:2010vc,Adams:2013qma} demonstrated the noise and SGWB discerning by employing three optimal channels from the Michelson.

For the first-generation TDI, besides the Michelson (X, Y, Z), there are four configurations, Sagnac ($\alpha, \beta, \gamma$), Relay (U, V, W), Beacon (P, Q, R), and Monitor (D, F, G) \cite{1999ApJ...527..814A,2000PhRvD..62d2002E,Tinto:2020fcc}.
Three optimal channels (A, E, T) could be composited from three regular channels \cite{Prince:2002hp,Vallisneri:2007xa}. The A and E channels will be sensitive to GWs and treated as science channels. The T channel is insensitive to GW and committed to instrument noises characterization. However, the T channel could be not enough to characterize all noise components, and additional data may be needed. For the science TDI channels, due to the worse GW response and severer acceleration noise in lower frequencies, the GW-insensitive data from the very low-frequency band may be utilized for acceleration noise characterization \cite{Caprini:2019pxz,Flauger:2020qyi}.

Much more second-generation TDI observables could be constructed by synthesizing the first-generation observables with different orders \cite{Vallisneri:2005ji}. These second-generation TDI observables, as well as the first-generation observables, could be decomposed into four first-generation generators ($\alpha$, $\beta$, $\gamma$ and $\zeta$) with different polynomial coefficients \cite{Dhurandhar:2002zcl,Hartwig:2021mzw}, and only three observable are expected to be independent. Even so, the different selections of triple TDI channels could yield different effects on noise characterization. Considering the second-generation TDIs are essentially derived from the first-generation observables, we limit the noise characterization in the first-generation channels in this work. 
Synergy is obtained by combining the Michelson and Relay configurations. The Michelson could have better capability to determine the acceleration noises than the Relay, and the Relay channels can break the degeneracy between the optical path noises and effectively resolve the noise parameters.

The investigations are performed by using three-year stationary Gaussian noise. The galactic foreground and a power-law astrophysical SGWB are simulated to evaluate the signal characterizing at the same time.
Their characterizations are fulfilled in two scenarios. In the first scenario, the noise components are characterized with signals-presented data, and the parameters of noises and signals are inferred simultaneously by combining the null-stream T channel(s) and science channels from the Michelson and/or Relay configurations. The second case is to characterize noise by selecting the GW-insensitive bands from science TDI channels and the T channel(s), and the parameters of noise components are estimated without considering the existence of a signal. With the characterized noises, the power spectral density (PSD) of instrumental noises in a TDI channel is restored for the targeting frequency band. Then by comparing the recovered noise PSD and observed data, the unmodelled GW signals may be recognized, especially for the frequencies in which the signal's PSD exceeds the noise. 

This paper is organized as follows. 
In Sec. \ref{sec:TDI_data}, we recap the first-generation TDI configurations and their noise components, and introduce data simulation with the galactic foreground and a power-law SGWB. 
In Sec. \ref{sec:char_noises}, we specify the algorithm for the noise characterization and analyze the results from different TDI combinations.
By using the characterized noises, we further reconstruct the spectral shapes of instrument noises and the injected signals in Sec. \ref{sec:reconstruction}.
We recapitulate our conclusions in Sec. \ref{sec:conclusions}. (We set $G=c=1$ in this work except otherwise stated).

\section{Time-delay interferometry and data simulation} \label{sec:TDI_data}

\subsection{Time-delay interferometry} \label{subsec:TDI}

\begin{figure}[htb]
\includegraphics[width=0.48\textwidth]{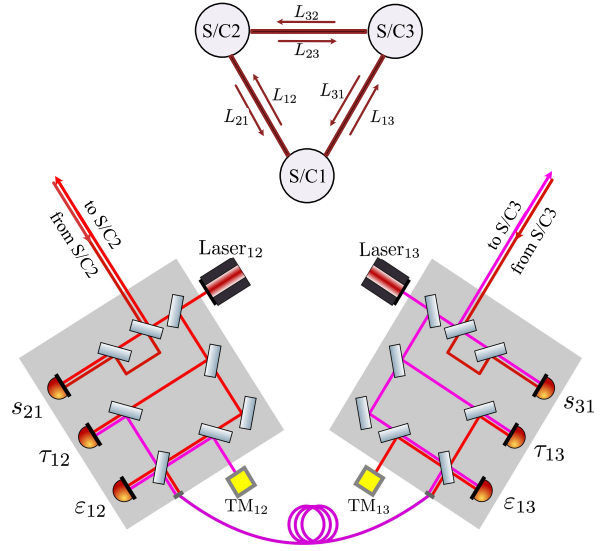} 
\caption{The triangular layout of three spacecraft and the diagram of two optical benches on S/C1 \cite{Otto:2015}. \label{fig:SC_layout} 
}
\end{figure}

TDI will be employed by LISA to suppress laser frequency noise by combining time-shifted interferometric links and forming equivalent equal paths. Five first-generation TDI configurations are developed based on the different topology paths, and three channels are included in each configuration depending on the different starting spacecraft (S/C) which are Michelson (X, Y, Z), Sagnac ($\alpha, \beta, \gamma$), Relay (U, V, W), Beacon (P, Q, R), and Monitor (D, F, G) \cite{1999ApJ...527..814A,2000PhRvD..62d2002E,Tinto:2020fcc}. The expressions of the first channels from five TDI configurations are
\begin{widetext}
\begin{align}
{\rm X} =& ( \mathcal{D}_{31} \mathcal{D}_{13} \mathcal{D}_{21} \eta_{12}  + \mathcal{D}_{31}  \mathcal{D}_{13} \eta_{21}  +  \mathcal{D}_{31} \eta_{13} +  \eta_{31}   ) - ( \eta_{21} + \mathcal{D}_{21} \eta_{12} +\mathcal{D}_{21} \mathcal{D}_{12} \eta_{31} + \mathcal{D}_{21}  \mathcal{D}_{12} \mathcal{D}_{31} \eta_{13} )  \label{eq:X_expression} \\
\alpha = & (\eta_{31} + \mathcal{D}_{31} \eta_{23} + \mathcal{D}_{31} \mathcal{D}_{23} \eta_{12} ) - ( \eta_{21} +   \mathcal{D}_{21} \eta_{32} + \mathcal{D}_{21} \mathcal{D}_{32} \eta_{13} ),  \label{eq:alpha_expression}  \\
{\rm U} =& (\eta_{23} + \mathcal{D}_{23} \eta_{32} + \mathcal{D}_{32}  \mathcal{D}_{23}  \eta_{13} + \mathcal{D}_{13} \mathcal{D}_{23}  \mathcal{D}_{32}  \eta_{21} )   - ( \eta_{13} + \mathcal{D}_{13} \eta_{21} + \mathcal{D}_{21}  \mathcal{D}_{13}  \eta_{32} + \mathcal{D}_{32} \mathcal{D}_{21}  \mathcal{D}_{13} \eta_{23}  ), \\
{\rm P} =& (\mathcal{D}_{13} \eta_{32} + \mathcal{D}_{13}  \mathcal{D}_{32} \eta_{23} + \mathcal{D}_{13}  \mathcal{D}_{32}  \mathcal{D}_{23} \eta_{12} + \mathcal{D}_{12} \eta_{13} ) - (  \mathcal{D}_{12}  \eta_{23} + \mathcal{D}_{12}  \mathcal{D}_{23}  \eta_{32} + \mathcal{D}_{12}                \mathcal{D}_{23}  \mathcal{D}_{32} \eta_{13} + \mathcal{D}_{13} \eta_{12} ), \\
{\rm D} =& (\eta_{21} + \mathcal{D}_{21}  \eta_{32} + \mathcal{D}_{21}  \mathcal{D}_{32} \eta_{23} + \mathcal{D}_{23} \mathcal{D}_{32} \eta_{31} )  - (  \eta_{31} + \mathcal{D}_{31}  \eta_{23} + \mathcal{D}_{31}  \mathcal{D}_{23}  \eta_{32} + \mathcal{D}_{23} \mathcal{D}_{32}  \eta_{21} ) \label{eq:D_expression},
\end{align}
\end{widetext}
where $\mathcal{D}_{ij}$ is a time-delay operator, $ \mathcal{D}_{ij} \eta(t) = \eta(t - L_{ij} )$, $L_{ij}$ is the arm length from S/C$i$ to $j$, $\eta_{ji}$ are Doppler measurement from S/C$j$ to S/C$i$ which is defined as follows \cite{Otto:2012dk,Otto:2015,Tinto:2018kij}, 
\begin{equation} \label{eq:eta}
\begin{aligned}
  \eta_{ji} &= s_{ji} + \frac{1}{2} \left[ \tau_{ij} - \varepsilon_{ij} + \mathcal{D}_{ji} ( 2 \tau_{ji} - \varepsilon_{ji} - \tau_{jk} ) \right] \\
  & \quad \mathrm{for} \  (2 \rightarrow 1), (3 \rightarrow 2) \ \mathrm{and} \ (1 \rightarrow 3), \\
  \eta_{ji} &= s_{ji} + \frac{1}{2} \left[ \tau_{ij} - \varepsilon_{ij}  + \mathcal{D}_{ji} ( \tau_{ji} - \varepsilon_{ji}    ) + \tau_{ik} -  \tau_{ij} \right] \\
   & \quad \mathrm{for} \  (1 \rightarrow 2), (2 \rightarrow 3)\ \mathrm{and}\ (3 \rightarrow 1).
\end{aligned}
\end{equation}
The deployments of two optical benches on S/C1 are illustrated in Fig. \ref{fig:SC_layout}. Three interferometer measurements, $s_{ji}$, $\varepsilon_{ij}$ and $\tau_{ij}$ for optical benches on S/C2 pointing to S/C1 (which denoted as  2$\rightarrow $1, 3$\rightarrow $2 and 1$\rightarrow $3) will be
\begin{equation} \label{eq:s_epsilon_tau_1}
\begin{aligned}
   s_{ji} = & y^h_{ji}:h + \mathcal{D}_{ji} C_{ji}(t) - C_{ij}(t) + n^{\rm op}_{ij}(t), \\
   \varepsilon_{ij} = & C_{ik}(t) - C_{ij}(t) + 2 n^{\rm acc}_{ij}(t) , \\
   \tau_{ij} = & C_{ik}(t) - C_{ij}(t) ,
\end{aligned}
\end{equation}
and measurements $s_{ij}$, $\varepsilon_{ij}$ and $\tau_{ij}$ for optical benches on 1$\rightarrow$2, 2$\rightarrow$3 and 3$\rightarrow$1 will be
\begin{equation} \label{eq:s_epsilon_tau_2}
\begin{aligned}
   s_{ji} &=  y^h_{ji}:h + \mathcal{D}_{ji} C_{ji}(t) - C_{ij}(t) + n^{\rm op}_{ij}(t), \\
   \varepsilon_{ij} &= C_{ik}(t) - C_{ij}(t) - 2 n^{\rm acc}_{ij}(t) , \\
   \tau_{ij} &= C_{ik}(t) - C_{ij}(t),
\end{aligned}
\end{equation}
where $y^h_{ji}$ is response function to the GW signal $h$ (see specific formula in Appendix \ref{sec:appendix_response}) \cite{1975GReGr...6..439E,1987GReGr..19.1101W,Vallisneri:2007xa,Vallisneri:2012np}, $C_{ij}$ denotes laser noise on the optical bench of S/C$i$ pointing to S/C$j$, $n^{\mathrm{op}}_{ij}$ represents the optical path noise on the S/C$i$ pointing to $j$, and $n^{\mathrm{acc}}_{ij}$ denotes the acceleration noise from test mass on the S/C$i$ pointing to $j$. In the following investigations, we assume the laser frequency noises are sufficiently suppressed, and the acceleration noises and optical path noises remain as the detection noises.

For the LISA with six laser links, a set of optimal channels (A, E, T) could be constructed from three regular channels (a, b, c) in a TDI configuration \cite{Prince:2002hp},
\begin{equation} \label{eq:optimalTDI}
 {\rm A} =  \frac{ {\rm c} - {\rm a} }{\sqrt{2}} , \quad {\rm E} = \frac{ {\rm a} - 2 {\rm b} + {\rm c} }{\sqrt{6}} , \quad {\rm T} = \frac{ {\rm a} + {\rm b} + {\rm c} }{\sqrt{3}}.
\end{equation}
The A and E channels, as two science data streams, could respond to GW effectively. The T channel is a null channel and insensitive to GW signals, and it could be utilized to characterize the detection noises. However, the T channel is dominated by the optical path noise, especially for equal-arm situations. On the other side, the acceleration noise overwhelms optical path noise for frequencies lower than 3 mHz in most regular TDI channels, and its characterization may require the low-frequency data from these channels.

\subsection{Data generation}  \label{subsec:data_generation}

The PSDs of acceleration noise and optical path noise for the LISA mission are targeted to be \cite{2017arXiv170200786A},
 \begin{align}
 S_{\rm acc} &= N^2_\mathrm{acc} \frac{\rm fm^2/s^4}{ \rm Hz } \left[ 1 + \left( \frac{0.4 \ {\rm mHz}}{f} \right)^2 \right]  
 \left[ 1 + \left(\frac{f}{8 \ {\rm mHz}} \right)^4 \right], \label{eq:Sn_acc} \\
 S_{\rm op} & =  N^2_\mathrm{op} \frac{\rm pm^2}{\rm Hz} \left[ 1 + \left(\frac{2 \ {\rm mHz}}{f} \right)^4 \right], \label{eq:Sn_op}
 \end{align}
where $N_\mathrm{acc} = 3$ and $N_\mathrm{op} = 10$ are the amplitudes of corresponding noise budgets. To distinguish the noises components on three spacecraft, $N_{ \mathrm{acc} ij}$ is labeled as the amplitude of acceleration noise associated with the optical bench on S/C$i$ pointing to S/C$j$, and $N_{ \mathrm{op} ij}$ is the amplitude of optical metrology noise from optical bench on S/C$i$ facing to S/C$j$. Considering each spacecraft carries two optical benches, twelve independent noise components are counted to evaluate the noise level of a TDI channel. The coefficients of the acceleration noise and optical path noise components for PSD and cross-spectral density (CSD) of typical TDI channels are listed in Table \ref{tab:TDI_acc} and Table \ref{tab:TDI_op}, respectively. The coefficients for a TDI channel from the same configuration could be deduced by shifting the indexes of the factors.

\begin{table*}[htb]
\caption{\label{tab:TDI_acc} The coefficients of the acceleration noise components for PSD ($S_{aa}$) and CSD ($S_{ab}$) of selected TDI channels ($x = 2 \pi f L$). }
\begin{ruledtabular}
\begin{tabular}{c|c|c|c|c|c|c}
& $S_{\mathrm{acc}12 }$ & $S_{\mathrm{acc}13 }$ & $S_{\mathrm{acc}21 }$ & $S_{\mathrm{acc}23 }$ & $S_{\mathrm{acc}31 }$ & $S_{\mathrm{acc}32 }$  \\
\hline
$S_\mathrm{XX} $ & $ 4 \sin^2 2 x   $  &  $ 4 \sin^2 2 x   $ & $ 16 \sin^2 x $ & 0 & $ 16 \sin^2 x $ & 0 \\ 
$S_\mathrm{XY} $ & $-16 \cos x \sin^2 x $ & 0 & $-16 \cos x \sin^2 x $ & 0 & 0 & 0 \\
\hline
$S_{\alpha \alpha}$ & $ 4 \sin^2 \frac{3x}{2}  $ & $ 4 \sin^2 \frac{3x}{2}  $ & $ 4 \sin^2 \frac{x}{2}  $ & $ 4 \sin^2 \frac{x}{2}  $ & $ 4 \sin^2 \frac{x}{2} $ & $ 4 \sin^2 \frac{x}{2} $ \\
$S_{\alpha \beta} $ & $ -4 (1 + 2 \cos x) \sin^2 \frac{x}{2} $ & $ 4 (1 + 2 \cos x) \sin^2 \frac{x}{2} $ & $ -4 (1 + 2 \cos x) \sin^2 \frac{x}{2} $ & $ 4 (1 + 2 \cos x) \sin^2 \frac{x}{2} $ & $ -4 \sin^2 \frac{x}{2} $ & $ -4 \sin^2 \frac{x}{2} $  \\
 \hline
 $ S_\mathrm{DD} $ &  $ 4 \sin^2 x  $ & $ 4 \sin^2 x $ &  $ 4 \sin^2 x $ & $ 8 - 4 \sin^2 x - 8 \cos x $ &  $ 4  \sin^2 x $ &  $ 8 - 4 \sin^2 x - 8 \cos x $ \\
$ S_\mathrm{DF} $ & $ -4 \cos x \sin^2 x $ & $32 \cos^2 \frac{x}{2} \sin^4 \frac{x}{2} $ & $-4 \cos x \sin^2 x $ & $32 \cos^2 \frac{x}{2} \sin^4 \frac{x}{2} $ & 0 & 0 \\
\hline
 $ S_\mathrm{PP} $ & 
$  4  \sin^2 x $ & $ 4  \sin^2 x $ & $ 4  \sin^2 x  $ & $ 8 - 4 \sin^2 x - 8 \cos x $ & $  4 \sin^2 x  $ & $ 8 - 4 \sin^2 x - 8 \cos x $ \\
$ S_\mathrm{PQ} $ & $ -4 \cos x \sin^2 x $ & $32 \cos^2 \frac{x}{2} \sin^4 \frac{x}{2} $ & $ -4 \cos x \sin^2 x $ & $32 \cos^2 \frac{x}{2} \sin^4 \frac{x}{2} $ & 0 & 0 \\
 \hline
 $ S_\mathrm{UU} $ & 
$ 4 \sin^2 x  $ & $ 4 \sin^2 x  $ & $ 4 \sin^2 x  $ & $ 8 - 4 \sin^2 x - 8 \cos x \cos 2 x $ & $ 4 \sin^2 x  $ & $ 8 - 4 \sin^2 x - 8 \cos x \cos 2 x $ \\
 $ S_\mathrm{UV} $ & $ 4 \cos x \sin^2 x $ & $32 \cos^2 \frac{x}{2} \sin^4 \frac{x}{2} $ & $ 4 \cos x \sin^2 x $ & $32 \cos^2 \frac{x}{2} \sin^4 \frac{x}{2} $ & $ -2 \sin^2 2 x $ & $ -2 \sin^2 2 x $ \\
 \end{tabular}
\end{ruledtabular}
\end{table*}

\begin{table*}[htb]
\caption{\label{tab:TDI_op} The coefficients of the optical path noise components for PSD ($S_{aa}$) and CSD ($S_{ab}$) of selected TDI channels ($x = 2 \pi f L$). }
\begin{ruledtabular}
\begin{tabular}{c|c|c|c|c|c|c}
& $S_{\mathrm{op}12 }$ & $S_{\mathrm{op}13 }$ & $S_{\mathrm{op}21 }$ & $S_{\mathrm{op}23 }$ & $S_{\mathrm{op}31 }$ & $S_{\mathrm{op}32 }$ \\
\hline
$S_\mathrm{XX} $ & $ 4 \sin^2 x  $ & $ 4 \sin^2 x $ & $ 4 \sin^2 x $ & 0 & $ 4  \sin^2 x $ & 0 \\
$S_\mathrm{XY} $ & $ -4 \cos x \sin^2 x $ & 0 & $ -4 \cos x \sin^2 x $ & 0 & 0 & 0 \\
\hline
$S_{\alpha \alpha} $ & 1 & 1 & 1 & 1 & 1 & 1 \\
$S_{\alpha \beta} $ & $ \cos 2 x $ & $ \cos x $ & $ \cos 2 x $ & $ \cos x $ & $ \cos x $ & $ \cos x $ \\
 \hline
$S_\mathrm{DD} $ & $ 4  \sin^2 x  $ & $ 4  \sin^2 x  $ & 0 & $ 4 \sin^2 \frac{x}{2}  $ &  0 & $ 4 \sin^2 \frac{x}{2}  $  \\
$S_\mathrm{DF} $ & 0 & $ 2 \sin^2 x $ & 0 & $ 2 \sin^2 x $ & 0 & 0 \\
\hline
$S_\mathrm{PP} $ & 0 &  0 & $ 4 \sin^2 x $ & $ 4 \sin^2 \frac{x}{2}  $ & $ 4 \sin^2 x $ & $ 4 \sin^2 \frac{x}{2}  $  \\
$S_\mathrm{PQ} $ & 0 & 0 & 0 & 0 & $ 2 \sin^2 x $ & $ 2 \sin^2 x $ \\
\hline
 $ S_\mathrm{UU} $ & 0 & $ 4 \sin^2 x $ & $ 4 \sin^2 x  $ & $ 4 \sin^2 \frac{3x}{2}  $ & 0 &  $ 4 \sin^2 \frac{x}{2}  $  \\
 $ S_\mathrm{UV} $ & 0 & $ -2 \sin^2 x $ & $ 4 \cos x \sin^2 x $ & 0 & 0 & $ -2 \sin^2 x $
\end{tabular}
\end{ruledtabular}
\end{table*}

The Gaussian noise streams in the time-domain are generated for twelve components based on the noise budgets in Eqs. \eqref{eq:Sn_acc} and \eqref{eq:Sn_op}, and these streams are time-shifted and synthesized to generate the output data of the TDI channels. The time-shifts are implemented by using Fourier transform and inverse Fourier transform under a static unequal-arm configuration, $L_{12} = L_{21} = 8.417 $ s, $L_{13} = L_{31} = 8.250 $ s and $L_{23} = L_{32} = 8.334 $ s. And the synthesizes are done by following Eqs. \eqref{eq:X_expression}-\eqref{eq:s_epsilon_tau_2}. The data of the optimal channels for each configuration are constructed by implementing Eqs. \eqref{eq:optimalTDI}. The sampling interval is set to be 25 s which yields a Nyquist frequency equal to 20 mHz. The reason for this high-frequency cutoff is that the T channel will become sensitive to GW signals for frequencies higher than $\sim$20 mHz \cite{Prince:2002hp,Vallisneri:2007xa,Wang:1stTDI}. On the other side, for the unequal arm case, the T channel could also be sensitive as the science channels at frequencies lower than $\sim$0.2 mHz \cite{Adams:2010vc,Wang:1stTDI}. Considering the poor GW response and severe acceleration noise, the SGWB may not be observed in this low frequency band, and the data could still be used for the noise characterization \cite{Caprini:2019pxz,Flauger:2020qyi}.

\begin{figure}[htb]
\includegraphics[width=0.48\textwidth]{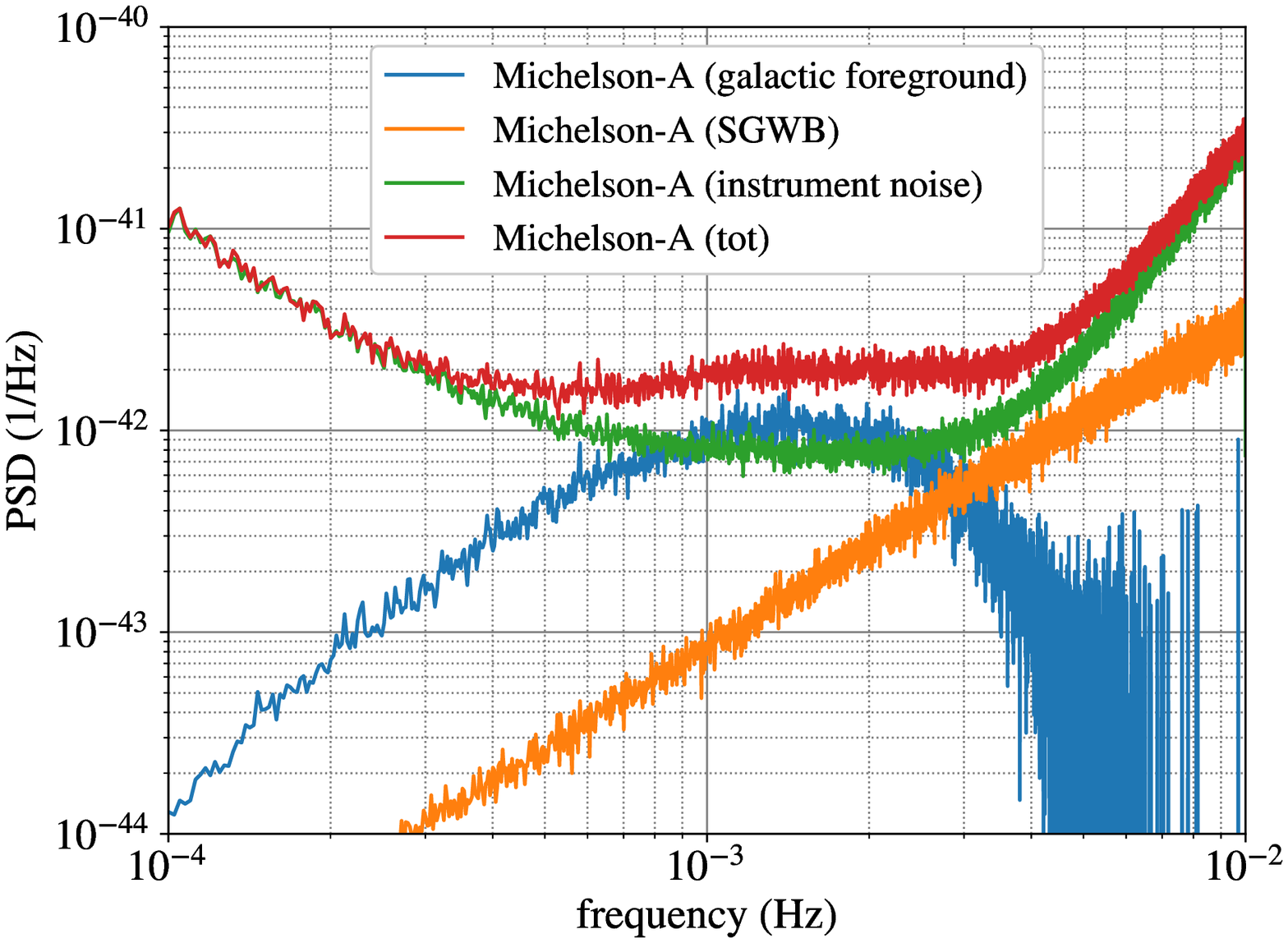}
\includegraphics[width=0.48\textwidth]{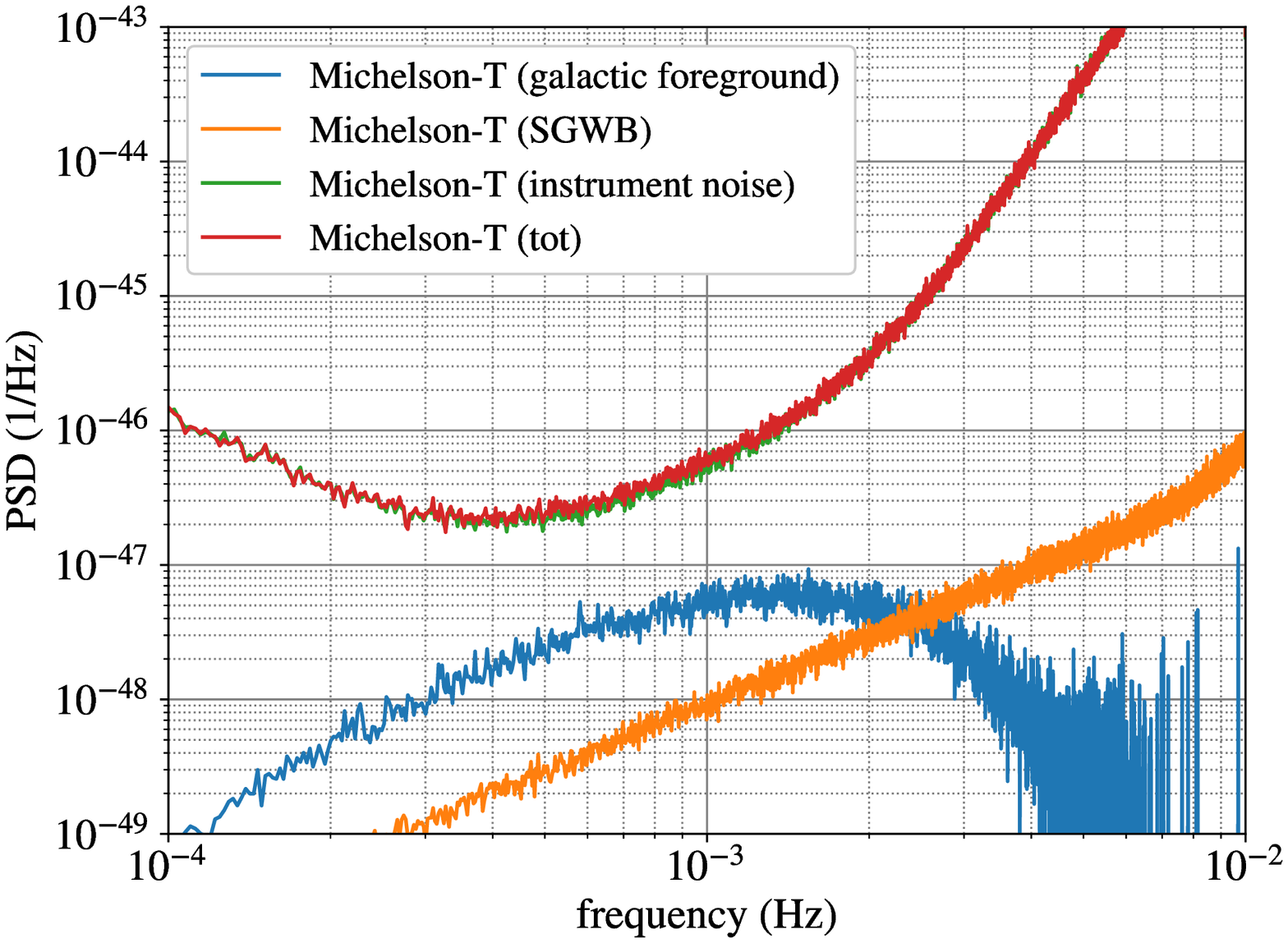}
\caption{\label{fig:LISA_AT_PSD} The PSDs of simulated instrument noise, galactic foreground and SGWB in the Michelson-A channel (upper panel) and T channel (lower panel). The A channel is sensitive to the injected GW signals in a frequency band of $\sim$[0.3, 6] mHz, and the null T channel is insensitive to the injected signals. The regular channels and optimal channels from other TDI configurations (Relay, Beacon, and Monitor) are expected to have similar detectability.
}
\end{figure}

Two GW signals are selected and injected into the data. The first one is galactic foreground, and the second one is a power-law isotropic SGWB. The galactic foreground is yielded from unresolved compact binaries in our galaxy. The population of these binary systems utilized in this work is from the LISA Data Challenge which includes $\sim$30 million binaries \cite{LDC}. The total PSD of the galactic foreground is approximated as follows which is modified from the formula in \cite{Cornish:2017vip},
\begin{equation}
P_\mathrm{GB} = A_\mathrm{GB} \times 10^{-45}  f^{-7/3} \left[ 1 + \tanh( \gamma ( f_k - f ) ) \right], \label{eq:GB_signal}
\end{equation}
where $A_\mathrm{GB} \simeq 1.4$, $\gamma \simeq 900$, and $f_k \simeq 1.29$ mHz. We clarify that the galactic foreground is an anisotropic signal due to the uneven distribution of the compact binaries in the Milky Way, and we ignore the amplitude modulation with LISA's yearly orbital motion in this work. 
 
The SGWB could be generated by both astrophysical sources and cosmological mechanisms with various spectral shapes in the LISA sensitive band \cite[reference therein]{Caprini:2015zlo,Caprini:2019egz,Kuroyanagi:2018csn}. In this investigation, a power-law SGWB is selected which could be produced by unresolved BH and NS binaries out of the Milky Way, and its PSD is expected to be \cite{LIGOScientific:2019vic,Caprini:2019pxz,Flauger:2020qyi,Martinovic:2020hru},
\begin{equation}
P_\mathrm{SGWB} = A_{0} \left( \frac{f}{f_\mathrm{ref}} \right)^{\alpha_0}  \frac{3 H^2_0}{4 \pi^2 f^3}, \label{eq:SGWB_signal} 
\end{equation}
where $H_0 \simeq 2.185 \times 10^{-18}$ Hz is the Hubble constant \cite{Planck:2018vyg}, $A_{0}$ is the amplitude of the SGWB energy density, and $\alpha_{0}$ is the index of the power law. The fiducial values for the power-law shape are chosen to be $A_0 = 4.446 \times 10^{-12}$ and $\alpha_0 = 2/3$ at reference frequency $f_\mathrm{ref} = 1 \ \mathrm{mHz} $ \cite{LIGOScientific:2019vic}.

Considering the antenna pattern of an interferometer, the PSD of an observed GW signal will be 
\begin{equation}
S_{h, \rm TDI} (f) = P_h (f)  \mathcal{R}_\mathrm{TDI} (f), \label{eq:S_h}
\end{equation}
where $ \mathcal{R}_\mathrm{TDI} $ is averaged response function of a TDI channel which is estimated by
\begin{equation} 
\begin{aligned}
 \mathcal{R}_{\rm TDI} (f) =& \frac{1}{4 \pi^2}  \int |F^h_{ \rm TDI} (f, \Omega)|^2 {\rm d} \Omega,
\end{aligned}
\end{equation}
and instantaneous response $F^h_{ \rm TDI}$ for each TDI channel is calculated by using Eqs. \eqref{eq:source_vec}-\eqref{eq:y_ij} and Eqs. \eqref{eq:X_expression}-\eqref{eq:D_expression}. Then the frequency-domain signals could be generated from their PSD shapes for a TDI channel \cite{Caprini:2015zlo},
\begin{equation}
 S^\mathrm{inj}_{h, \mathrm{TDI} } ( f ) = \frac{1}{2} \bigg| \mathcal{G} (0,  S^{1/2}_{h,\mathrm{TDI}} ( f )  ) + i \mathcal{G} ( 0,  S^{1/2}_{h,\mathrm{TDI}} ( f )  ) \bigg|^2,
\end{equation}
where $\mathcal{G}( \mu, \sigma)$ is a random number generator following Gaussian distribution with mean value $\mu$ and standard deviation $\sigma$.

The LISA is designed to be a 4 years mission and would be extensible for up to 10 years \cite{2017arXiv170200786A}. However, only a 75\% scientific duty cycle could be expected because of the interruption by antenna repositioning and other operations \cite{Caprini:2019pxz}.  And only 3 years of data will be effective in a 4 years observation. For the stochastic signals and stationary noise, the impact of data discontinuity is expected to be insignificant. Therefore, the data is generated for 3 years continuously. The PSD and cross spectral density (CSD) of the time-domain noise data are calculated for selected TDI channels, and the simulated GW signals are injected into the noise data by applying
\begin{equation}
 S_{\mathrm{data}} (f) = S_{\mathrm{n, inst} } ( f ) +  S^\mathrm{inj}_{h, \mathrm{SGWB} } ( f ) +  S^\mathrm{inj}_{h, \mathrm{GB} } ( f ) .
\end{equation}

The PSDs of instrument noise and simulated GW signals for Michelson-A and T channels are illustrated in Fig. \ref{fig:LISA_AT_PSD}. As we can see, the A channels will be sensitive to the injected GW signals in the frequency band of $\sim$[0.3, 6] mHz. Similar to the Michelson-A, the other science TDI channels are also expected to be sensitive to injected signals around the same frequency band. The Michelson-T, as a null stream, is insensitive to the injected signals. The targeting frequency band for the noise characterization is selected to be within a frequency range of [0.02, 20] mHz. The low-frequency cutoff is close to the low boundary of the LISA observational band, and the high-frequency cutoff is set to keep the T channel insensitive to GW signals as aforementioned.

\section{Characterizing instrumental noises from TDI combinations} \label{sec:char_noises}

In this section, with the simulated data, we characterize the noises by inferring the parameters of noise components from TDI channel combinations.

\subsection{Algorithm for parameter inference}

The Bayesian inference based on the Markov chain Monte Carlo is employed to estimate the parameters of noises or signals. Although all first-generation TDI observables could be produced from four generators ($\alpha$, $\beta$, $\gamma$, and $\zeta$) \cite{Dhurandhar:2002zcl,Tinto:2020fcc}, the different TDI channels could yield different performances on noises and signals characterization, and the Relay configuration is chosen to explicate the reason in Appendix \ref{sec:Sagnac_Relay}.
As Table \ref{tab:TDI_acc} show, except for Sagnac-$\alpha$ channels, the coefficients of paired acceleration noise components, $S_{\mathrm{acc}ij}$ and $S_{\mathrm{acc}ji}$, are (closely) equal for each TDI channel at low frequencies. For the Michelson-X channel, since the factor of $\cos^2 x_{ij}$ approaches 1 when $x_{ij}$ is small, the coefficients of the paired acceleration noises are closely equal at the low frequency. This degeneracy also exists in the CSD of two channels from a TDI configuration. Although the Sagnac observables have asymmetric coefficients for the acceleration noise components, their acceleration noise is much lower than the optical path noise and could be hard to be characterized. 

Similarly, for the Sagnac and Michelson, their coefficients of optical path noises pairs, $S_{\mathrm{op}ij}$ and $S_{\mathrm{op}ji} $, are also identical for each TDI channel as listed in Table \ref{tab:TDI_op}. The equal coefficients indicate the degeneracy between these noise components, the best-measured values will be $ S_{\mathrm{op}ij} + S_{\mathrm{op}ji}$. With different coefficients, the Relay observables could break the degeneracy between the paired noise components, and they could be an efficient choice to combine with the Michelson for noise characterization. As we also verified, the joint two of three configurations, Michelson, Beacon, and Monitor, could also resolve optical metrology noises. 
Once the degeneracy between the optical path noises is resolved, their parameters could be determined with better accuracy from the combination. Considering the Relay observables could determine the optical path noise independently (or, in other words, without cross-correlation with other TDI configurations), the Relay are elected to combine with the Michelson for noise characterization and to compare the performances from individual and combined TDI configurations.

Two scenarios are implemented to characterize noises. The first one is noise-only combinations which determine the noise parameters by combining the T channel(s) and low-frequency data of A and E channels from Michelson and/or Relay configurations. The T channel is a null stream or GW-insensitive, and its data in the frequency range [0.02, 20] mHz are utilized. For the science TDI channels, by assuming they are GW-insensitive at very lower frequencies, their data in frequency band [0.02, 0.2] mHz are cautiously selected which is dominated by acceleration noises.
The second scenario is the science case, the parameters of noises and signals are inferred from three optimal channels from Michelson and/or Relay. The difference from the noise-only case is that the frequency range of A and E channels is extended from [0.02, 0.2] mHz to [0.02, 20] mHz. These data streams in a larger frequency band, as the science data, are employed to determine parameters of both noises and signals. 
The channels and frequency band selections for two scenarios are listed in Table \ref{tab:run_checklist}. For each scenario, three combinations are examined to compare their performances on parameters characterization.
\begin{table}[tbh]
\caption{\label{tab:run_checklist} The checklists of TDI combinations for channels and frequency bands selections. The combinations in noise-only case select the T channel(s) in a frequency band of [0.02, 20] mHz with/without the science TDI channels in a low-frequency band of [0.02, 0.2] mHz, and combinations in the science case include optimal channels in the frequency band of [0.02, 20] mHz.
}
\begin{ruledtabular}
\begin{tabular}{cccccc}
scenario &  combination & Michelson & Michelson &   Relay &  Relay  \\
 & & (A, E) & T & (A, E) & T \\
 &  &  (mHz) & (mHz)  & (mHz) & (mHz)   \\
\hline
 & Michelson-T & -  & [0.02, 20]  & - &  - \\
noise-only & Michelson & [0.02, 0.2]  & [0.02, 20]  & - & - \\
& Michelson & [0.02, 0.2]  & [0.02, 20]  &  [0.02, 0.2]  & [0.02, 20] \\
& + Relay & & & &  \\
\hline
 & Michelson & [0.02, 20]   & [0.02, 20]  & - & - \\
science & Relay & - & - & [0.02, 20]  & [0.02, 20]    \\
& Michelson & [0.02, 20]  & [0.02, 20]  &  [0.02, 20]  & [0.02, 20] \\
& + Relay & & & &  \\
\end{tabular}
\end{ruledtabular}
\end{table}

The likelihood function of parameter inferences will be \cite{Adams:2010vc},
\begin{equation}
\ln \mathcal{L} ( \vec{ \theta }, A_0, \alpha_0, A_\mathrm{GB} ) \propto - \frac{1}{2} \sum_i \left[  \tilde{ \mathbf{s} }^{\dagger} (f_i) \mathbf{\Sigma}^{-1} (f_i)  \tilde{ \mathbf{s} } (f_i) + \ln \det \mathbf{\Sigma} \right] ,
\end{equation}
where $\mathbf{\Sigma}$ is the correlation matrix of the selected TDI channels, $ \tilde{ \mathbf{s}}$ is data vector of these channels, $\vec{ \theta }$ represents twelve amplitude squares of acceleration noises $N^2_{\mathrm{acc}ij}$ and optical path noises $N^2_{\mathrm{op}ij} (i,j=1,2,3 \ \mathrm{and}\ i \neq j)$ in Eq. \eqref{eq:Sn_acc} and \eqref{eq:Sn_op}, $A_0$ and $\alpha_0$ are the amplitude and power index of the SGWB in Eq. \eqref{eq:SGWB_signal}, and $A_\mathrm{GB}$ is the amplitude of galactic foreground in Eq. \eqref{eq:GB_signal}.
The priors for parameters are set to be uniform in their selected ranges, $N^2_{\mathrm{acc}ij} \in [0, 30]$, $N^2_{\mathrm{op}ij} \in [0, 400]$, $A_0 \in [0, 8]$, $\alpha_0 \in [0, 2]$ and $A_\mathrm{GB} \in [0, 3]$. The respective posterior probabilities for the noise-only and science cases will be
\begin{align}
p ( \vec{ \theta } )  \propto & \pi_N( \vec{ \theta } ) \mathcal{L} ( \vec{ \theta } ) \\
p ( \vec{ \theta }, A_0, \alpha_0, A_\mathrm{GB} )  \propto & \pi_N( \vec{ \theta } ) \pi_S( A_0, \alpha_0, A_\mathrm{GB} ) \notag \\  & \times \mathcal{L} ( \vec{ \theta }, A_0, \alpha_0, A_\mathrm{GB} ) , 
\end{align}
where $\pi_N( \vec{ \theta } )$ is the prior distribution for noise amplitudes, $\pi_S( A_0, \alpha_0, A_\mathrm{GB} )$ is the prior distribution for parameters of signals. The MCMC sampler in \textsf{emcee} is utilized to run the Bayesian inference \cite{emcee}.

\subsection{Noises characterization from noise-only case}

\begin{figure*}[htb]
\includegraphics[width=0.49\textwidth]{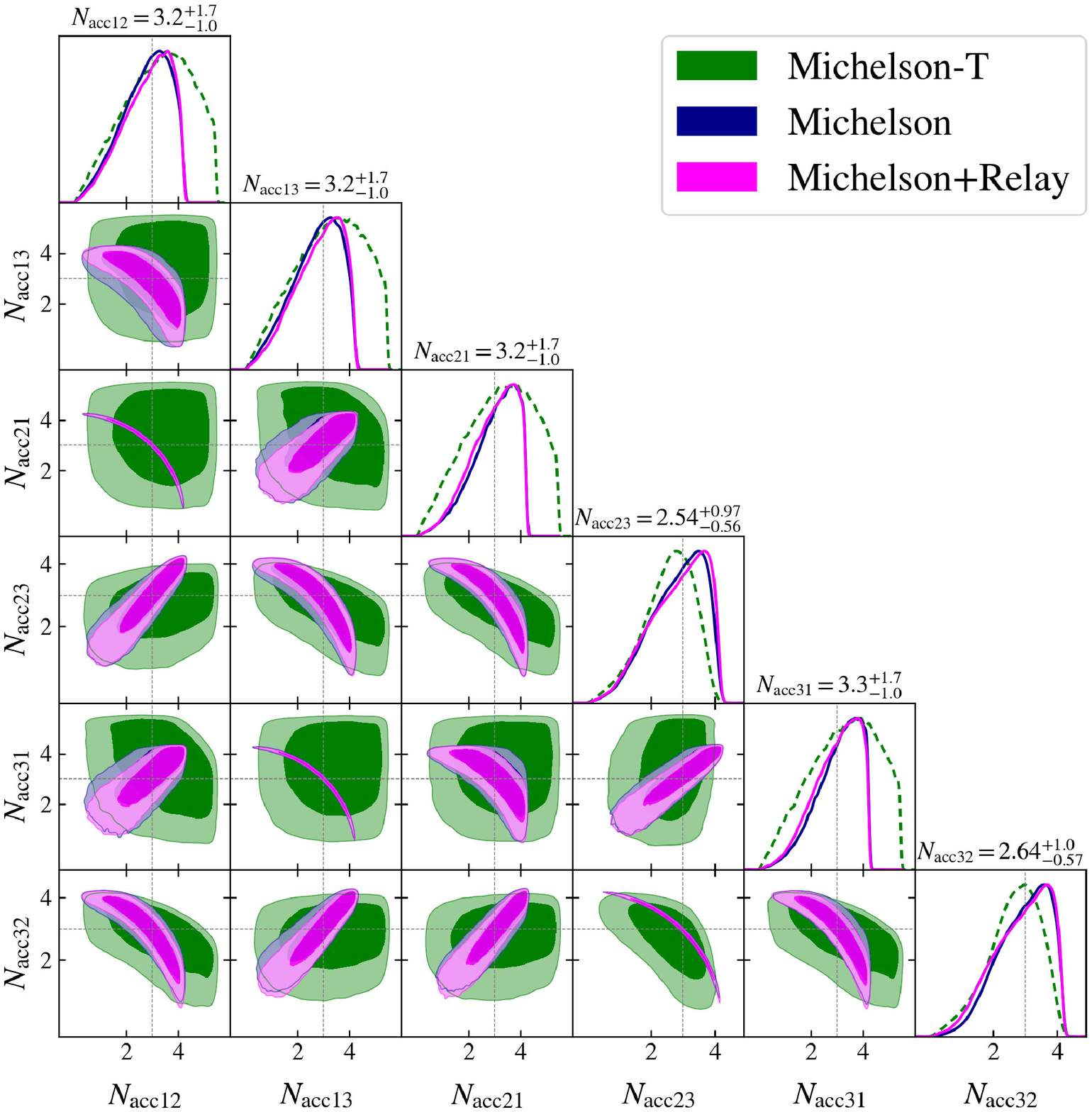} 
\includegraphics[width=0.49\textwidth]{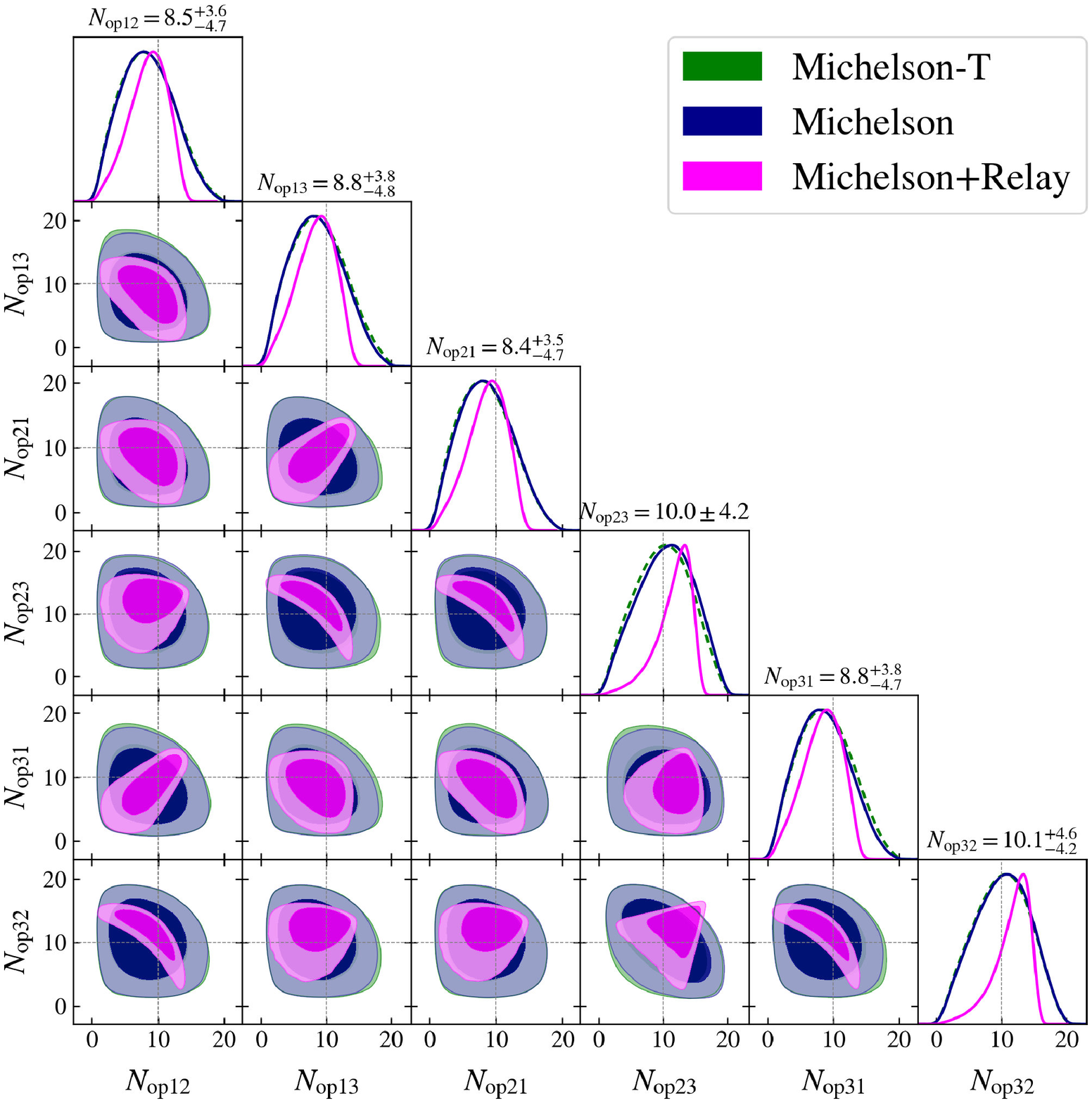} 
\caption{The corner plots for the amplitudes of acceleration noises (left panel) and optical path noises (right panel) inferred from three noise-only combinations which are 1) Michelson-T: the T channel from Michelson configuration in the frequency band [0.02, 20] mHz, 2) Michelson: combining the Michelson-T channel data in a frequency band [0.02, 20] mHz and Michelson-A and E channels in a band of [0.02, 0.2] mHz, and 3) Michelson+Relay combination: combining data of the Michelson-T and Relay-T channels in [0.02, 20] mHz and the low-frequency data of A and E channels from Michelson and Relay in [0.02, 0.2] mHz. The $1\sigma$ uncertainties from the Michelson-T channel are shown on the top of each column. In the left plot, the blue curves and magenta curves are overlapped which indicates acceleration noise characterizations from Michelson+Relay and Michelson are identical. In the right plot, green curves and blues curves are overlapped since low-frequency data from Michelson-A and E channels do not yield additional constraints on the optical path noises. \label{fig:corner_acc_op_noise_only} 
}
\end{figure*}

The noise-only combinations dedicate to acceleration noises and optical path noise characterization by using GW-insensitive datasets. 
To compare the accuracies of characterization from different datasets, three combinations are examined which include the single Michelson-T channel, Michelson and Michelson+Relay combinations as listed in Table \ref{tab:run_checklist}. The Michelson-T is a fiducial null stream to characterize the noises, and the results are shown by the green curves in Fig. \ref{fig:corner_acc_op_noise_only}. The $1 \sigma$ uncertainties of the noise parameters from the Michelson-T are listed on top of each column. The Michelson case incorporates the Michelson-T channel and low frequency band of science channels, and the results are shown by the blue curves in Fig. \ref{fig:corner_acc_op_noise_only}. The last combinator is the Michelson+Relay which combines two T channels from Michelson with Relay and low-frequency data of their four science channels, Michelson (A, E) and Relay (A, E), and the results are shown by the magenta curves.

The left panel of Fig. \ref{fig:corner_acc_op_noise_only} shows the amplitude characterization for six acceleration noise components. Compared to the results from the single Michelson-T channel, the Michelson combination could reduce the uncertainties of amplitude determinations, and this improvement is contributed by the additional low-frequency data from Michelson-A and E channels. On the other side, as the overlapped blue and magenta curves are shown, the Michelson+Relay combination does not show improvement compared to the Michelson. We suppose no additional information about acceleration noises from Relay channels contributes to the Michelson result. For both Michelson and Michelson+Relay combination, there are degeneracies between three pairs of noise components, $(N_{\mathrm{acc}ij}, N_{\mathrm{acc}ji} )$, as we expected. The degenerated acceleration noises are from two test masses in a laser link between two spacecraft. We also can notice that Michelson-T could constrain the $N_{\mathrm{acc}23}$ and $N_{\mathrm{acc}32}$ in a sensible range, and this should be attributed to the unequal arms. For real mission operations, the arm lengths between spacecraft will be different and vary with time, then the capability of the T channel could be enhanced for noise characterization. 

The characterizations of optical path noise are shown in the right panel of Fig. \ref{fig:corner_acc_op_noise_only}. The results from the Michelson-T and Michelson cases are largely overlapped because little information about optical path noise could be extracted from acceleration noise dominated Michelson data streams. When the Relay data streams are combined, the optical metrology noises can be characterized with better precision since the additional data streams are more sensitive to these noise components than the Michelson.

\subsection{Signals and noises characterization from science case}

\begin{figure*}[htb]
\includegraphics[width=0.49\textwidth]{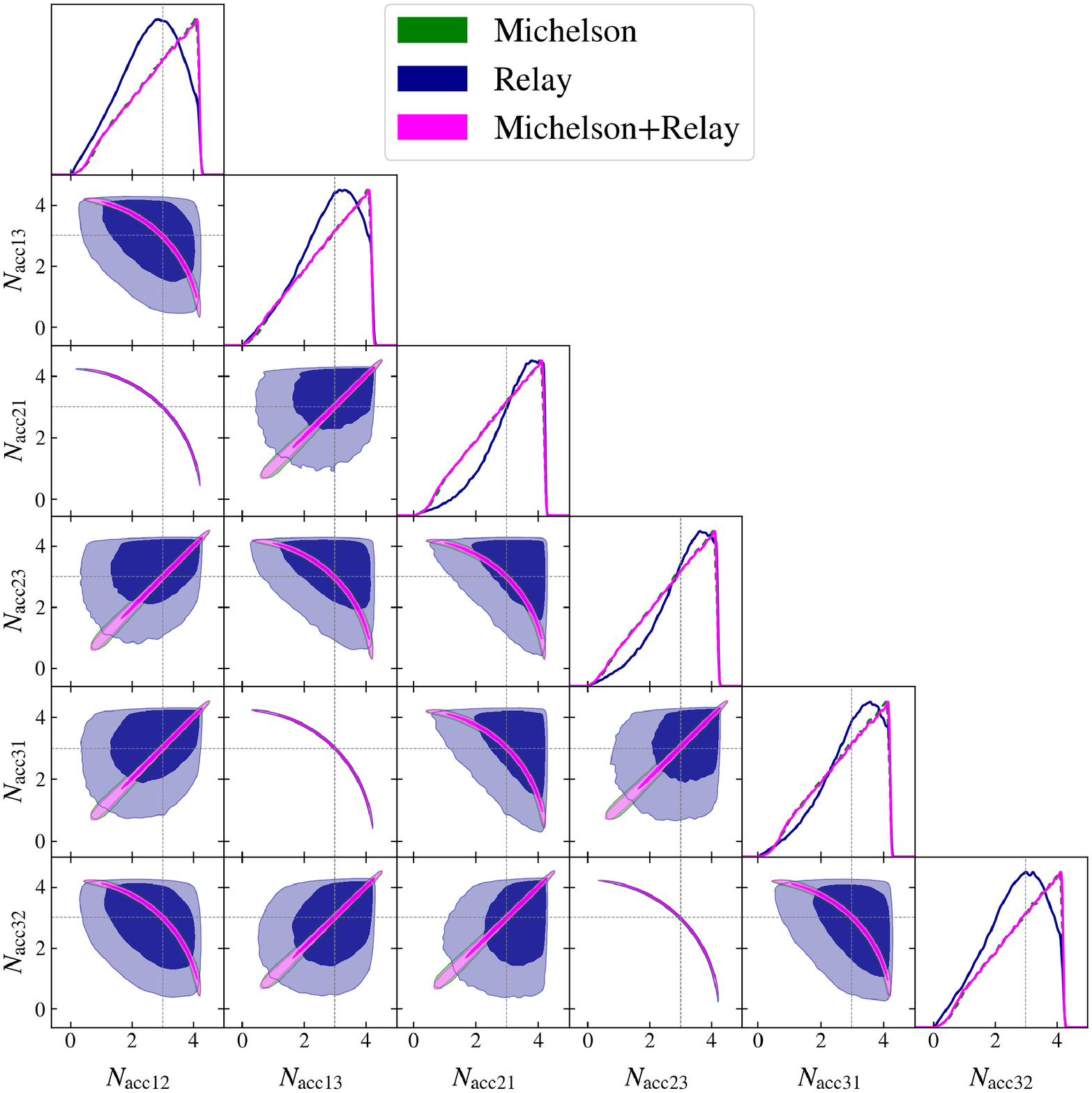} 
\includegraphics[width=0.49\textwidth]{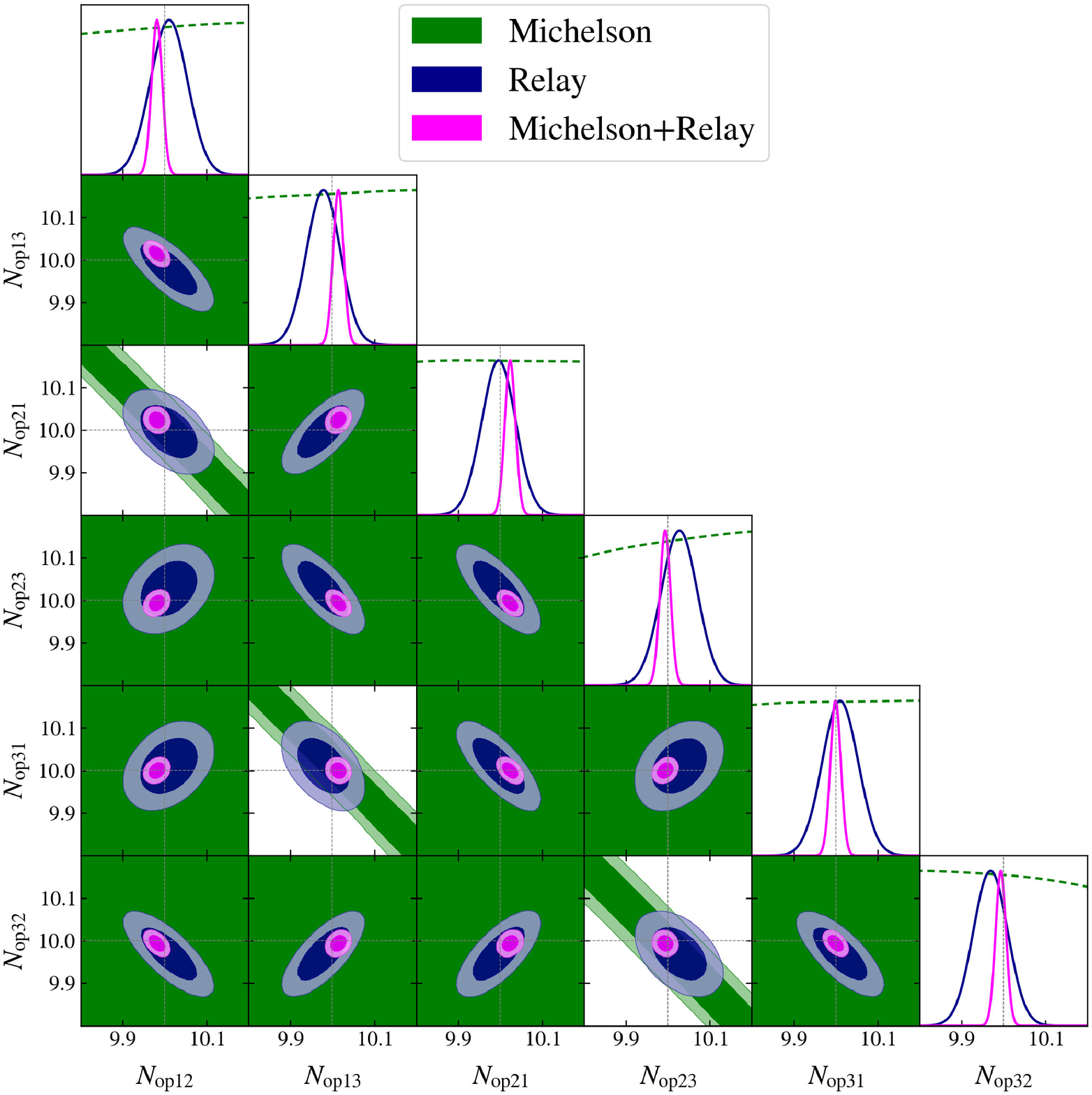} 
\includegraphics[width=0.32\textwidth]{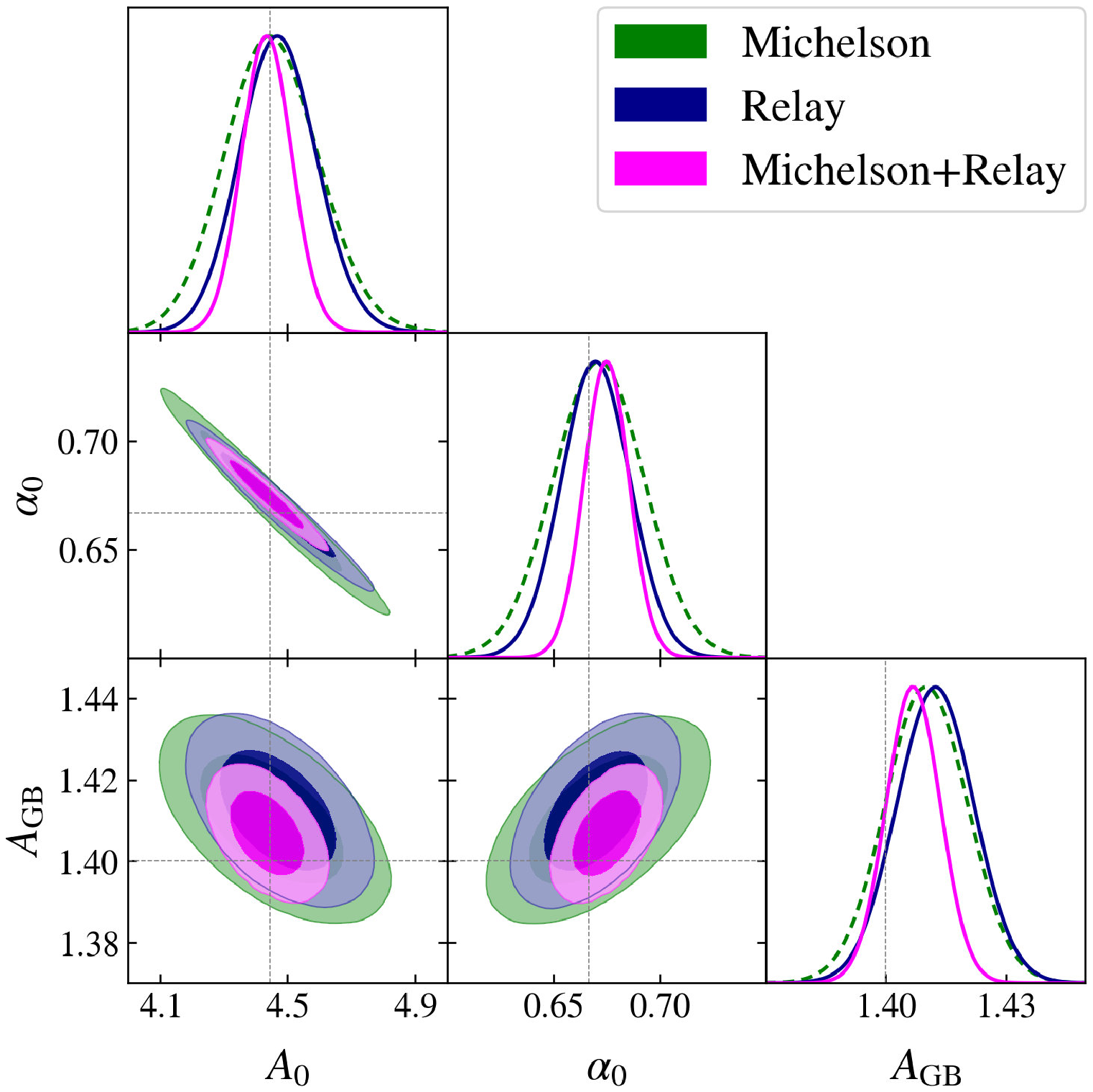}
\caption{The corner plots for the amplitudes of acceleration noises (upper left panel), optical path noises on six optical benches (upper right panel) and GW signals (lower panel) inferred from different TDI combinations which are 1) Michelson: combining three optimal channels (A, E, T) channels in a frequency band of [0.02, 20] mHz, 2) Relay: combining optimal channels of Relay in a frequency band of [0.02, 20] mHz, and 3) Michelson+Relay combination: combining six optimal channels from Michelson and Relay. The $1 \sigma$ uncertainties of these parameters are listed in Table \ref{tab:mcmc_result}. In the upper left plot, the green curves and magenta curves are overlapped which means no additional improvement is contributed from Relay data streams for acceleration noise determination. In the upper right plot, the areas and curves from the Michelson (green) are cropped to make the results from the Relay and Michelson+Relay more visible. \label{fig:corner_acc_op_signal} 
}
\end{figure*}

The science case characterizes both noises and signals from the TDI combinations. Compared to the noise-only case, the combinators in the science case employ additional GW-sensitive data from Michelson and Relay regular channels. Three combinations are implemented which are 1) Michelson which combines the optimal channels of Michelson in the frequency band of [0.02, 20] mHz, 2) Relay which includes optimal channels of Relay configuration in the band of [0.02, 20] mHz, and 3) Michelson+Relay which unifies the six channels from Michelson and Relay combinations.

\begin{figure}[htb]
\includegraphics[width=0.48\textwidth]{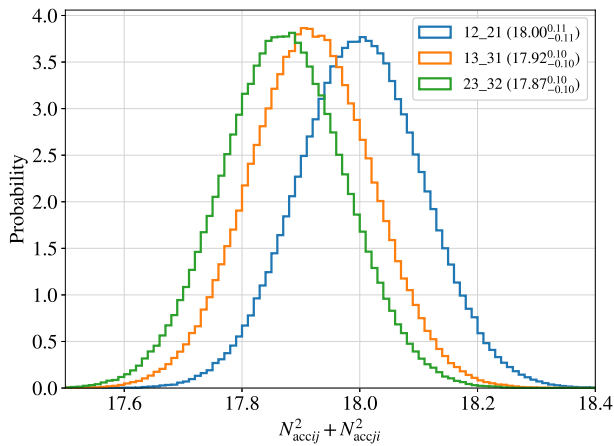} 
\caption{The histograms of the combined amplitudes between two degenerated acceleration noise components. The legend $ij\_ji$ indicates the correspond amplitude pair, for instance, $12\_21$ means the value from $N^2_{\mathrm{acc}12} + N^2_{\mathrm{acc}21}$. \label{fig:degeneracy} 
}
\end{figure}

The results from three combinations are shown in Fig. \ref{fig:corner_acc_op_signal}, and the inferred values with $1 \sigma$ uncertainties are listed in Table \ref{tab:mcmc_result}. The estimated amplitudes of the acceleration noises are shown in the upper left panel of Fig. \ref{fig:corner_acc_op_signal}. As we can see in the plots, for the Michelson and Michelson+Relay, their results are consistent which means no additional improvement is obtained from the Relay data. On the other side, the combined amplitudes, $N^2_{\mathrm{acc}ij} + N^2_{\mathrm{acc}ji}$, is better constrained compared to the results in the noise-only case, and this improvement should be obtained from the extra data of the science data streams. The histograms of three values are shown in Fig. \ref{fig:degeneracy}. 

\begin{table}[tbh]
\caption{\label{tab:mcmc_result} The $1\sigma$ uncertainties of parameters from three TDI combinations in the science case.
}
\begin{ruledtabular}
\begin{tabular}{cccc}
 &  &  &  Michelson   \\
parameter & Michelson  & Relay & +Relay \\
\hline
$N_\mathrm{acc12}$  & $ 3.004_{-1.234}^{0.854} $  & $ 2.662_{-1.111}^{0.919} $ & $ 3.002_{-1.249}^{0.862} $  \\
$N_\mathrm{acc13}$  & $ 2.977_{-1.245}^{0.863} $  & $ 2.886_{-1.110}^{0.836} $  & $ 2.978_{-1.261}^{0.870} $   \\
$N_\mathrm{acc21}$  & $ 2.995_{-1.230}^{0.860} $  & $ 3.302_{-1.027}^{0.645} $ & $ 2.999_{-1.246}^{0.865} $   \\
$N_\mathrm{acc23}$  & $ 2.981_{-1.251}^{0.859} $ & $ 3.200_{-1.059}^{0.693} $ & $ 2.977_{-1.264}^{0.869} $  \\
$N_\mathrm{acc31}$  & $ 3.002_{-1.234}^{0.855} $ & $ 3.090_{-1.085}^{0.747} $  & $ 3.008_{-1.244}^{0.861} $   \\
$N_\mathrm{acc32}$  & $ 2.996_{-1.229}^{0.860} $  & $ 2.761_{-1.117}^{0.883} $  & $ 3.000_{-1.247}^{0.864} $   \\
\hline
$N_\mathrm{op12}$  & $ 10.272_{-1.687}^{1.443} $ & $ 10.012_{-0.044}^{0.044} $  & $ 9.982_{-0.013}^{0.013}$   \\
$N_\mathrm{op13}$  & $ 10.134_{-1.696}^{1.451} $ & $ 9.977_{-0.040}^{0.040} $  & $ 10.014_{-0.012}^{0.012} $    \\
$N_\mathrm{op21}$  & $   9.732_{-1.796}^{1.516} $ & $ 9.994_{-0.040}^{0.040} $  & $ 10.024_{-0.013}^{0.013} $   \\
$N_\mathrm{op23}$  & $ 10.201_{-0.811}^{0.749} $ & $ 10.029_{-0.043}^{0.042} $  & $ 9.993_{-0.013}^{0.013} $   \\
$N_\mathrm{op31}$  & $   9.882_{-1.749}^{1.482} $ & $ 10.013_{-0.042}^{0.042} $  & $ 10.000_{-0.012}^{0.013} $   \\
$N_\mathrm{op32}$  & $   9.784_{-0.846}^{0.781} $ & $ 9.967_{-0.041}^{0.041} $ & $ 9.994_{-0.013}^{0.013} $    \\
\hline
$A_0$ & $ 4.431_{-0.143}^{0.148} $ & $ 4.471_{-0.118}^{0.121} $ & $ 4.436_{-0.076}^{0.078} $ \\
$\alpha_0$ & $ 0.673_{-0.021}^{0.021} $ & $ 0.670_{-0.016}^{0.016} $ & $ 0.675_{-0.011}^{0.011} $ \\
$A_\mathrm{GB}$ & $ 1.410_{-0.010}^{0.010} $ & $ 1.408_{-0.010}^{0.010} $ & $ 1.407_{-0.007}^{0.007} $
\end{tabular}
\end{ruledtabular}
\end{table}

The characterizations of optical metrology noises on six optical benches are shown in the upper right panel of Fig. \ref{fig:corner_acc_op_signal}.
The results from the Michelson configuration are shown by the green curves and areas. Because the uncertainties are much larger than the results from the Relay and Michelson+Relay as listed in Table \ref{tab:mcmc_result}, the curves of the Michelson are cropped to highlight the results from other two combinations. As we can partly read from the plot, the Michelson combination poorly resolves the optical path noises because of the degeneracies, and the best-determined values are $N^2_{\mathrm{op}ij} + N^2_{\mathrm{op}ji}$. Because degeneracy between optical path noise components could be resolved by Relay channels, the Relay combination can determine the amplitudes of optical noises with uncertainties of $\sim$0.8\%. For the Michelson+Relay case, the amplitudes of optical path noise can be precisely determined with uncertainties of $\sim$0.25\% which is essentially contributed by the Relay observables.

The determinations for signal parameters from three combinators are shown in the lower plot of Fig. \ref{fig:corner_acc_op_signal}. The results from the Relay combination are slightly better than Michelson because of better separated the power-law SGWB from resolved optical path noise in the higher frequency band. And their Michelson+Relay combination resolves three parameters with better accuracy. For the power-law SGWB, there is a degeneracy between parameters $A_0$ and $\alpha_0$ which will affect its PSD reconstruction in the next section.

In summary, the Relay data streams have much better capability than the Michelson to resolve the optical metrology noises, and the Michelson observables may constrain acceleration noises in a smaller parameter space than the Relay. A synergy for noise characterization could be achieved by combining these two configurations. Compared to the results from the Michelson+Relay combination, three observables from the Michelson could \textit{not} break the degeneracy between optical path noises as performed in \cite{Adams:2010vc}. If the noise components are assumed to be fully identical, the parameters of acceleration noise and optical path noise could be precisely determined from the Michelson data streams \cite{Caprini:2019pxz,Flauger:2020qyi,Boileau:2020rpg}. However, by ignoring the degeneracy between the noise components, the precision of noise parameters could be overestimated.

\section{Reconstructing spectral shapes of noise and signals} \label{sec:reconstruction}

In this section, the spectral shapes of the noises and signals in a TDI channel are reconstructed by using the results of the Michelson+Relay combination from the science and the noise-only cases, respectively. 
In the science case, the parameters of both noise components and signals are estimated. Therefore, the PSD of either noise or signal could be directly restored from the inferred values.
To calculate the confidence intervals of the recovered spectra, 5000 data arrays are randomly picked from achieved MCMC samples. Fifteen inferred parameters are included in each array, and twelve values of noise components are employed to calculate the noise PSD in a TDI channel. The values of $A_0$ and $\alpha_0$ are utilized to rebuild the spectral shape of SGWB by using Eq. \eqref{eq:SGWB_signal}, and $A_\mathrm{GB}$ is used to calculate the galactic foreground by using Eq. \eqref{eq:GB_signal}. 

\begin{figure}[htb]
\includegraphics[width=0.46\textwidth]{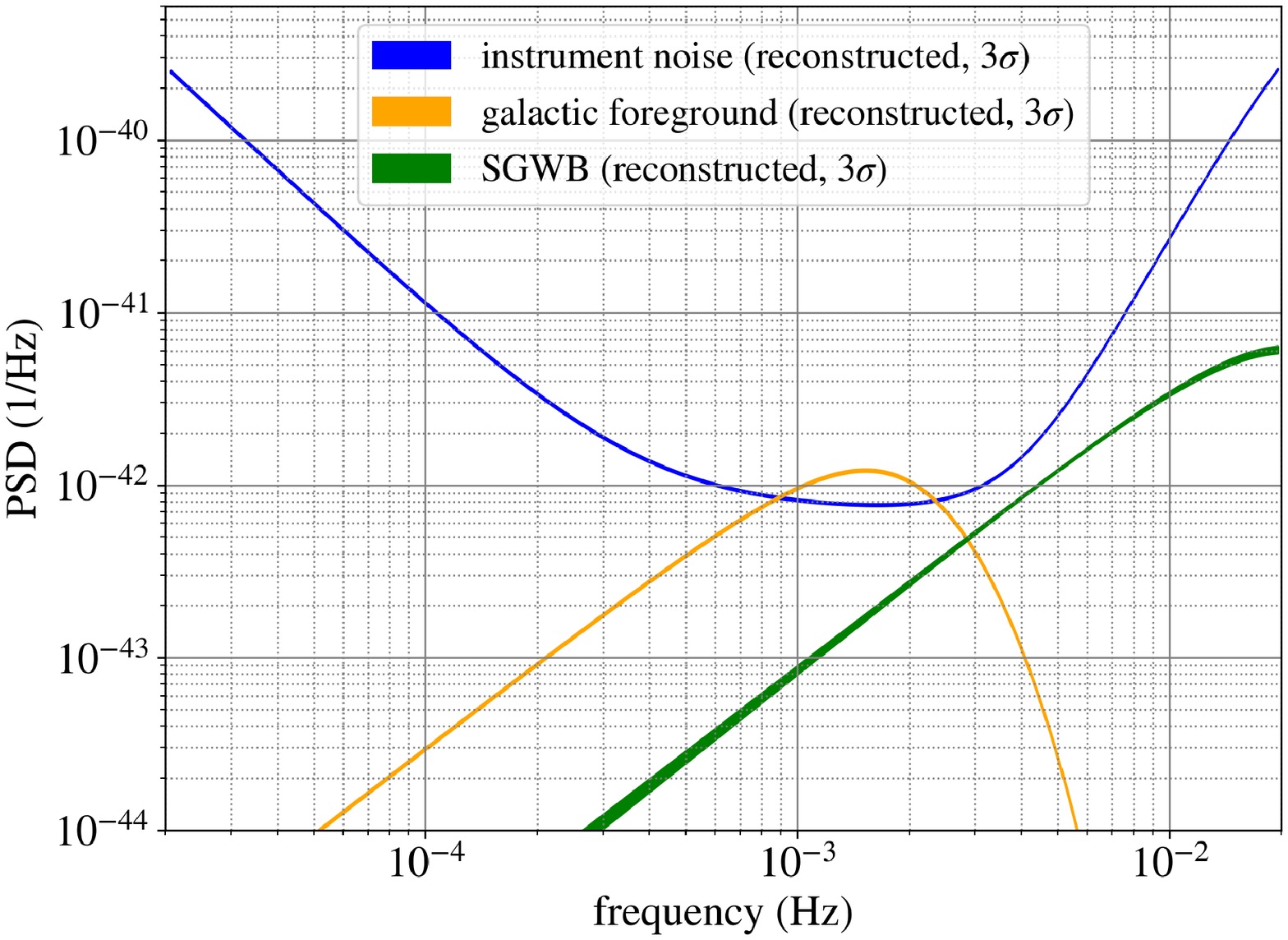} 
\includegraphics[width=0.46\textwidth]{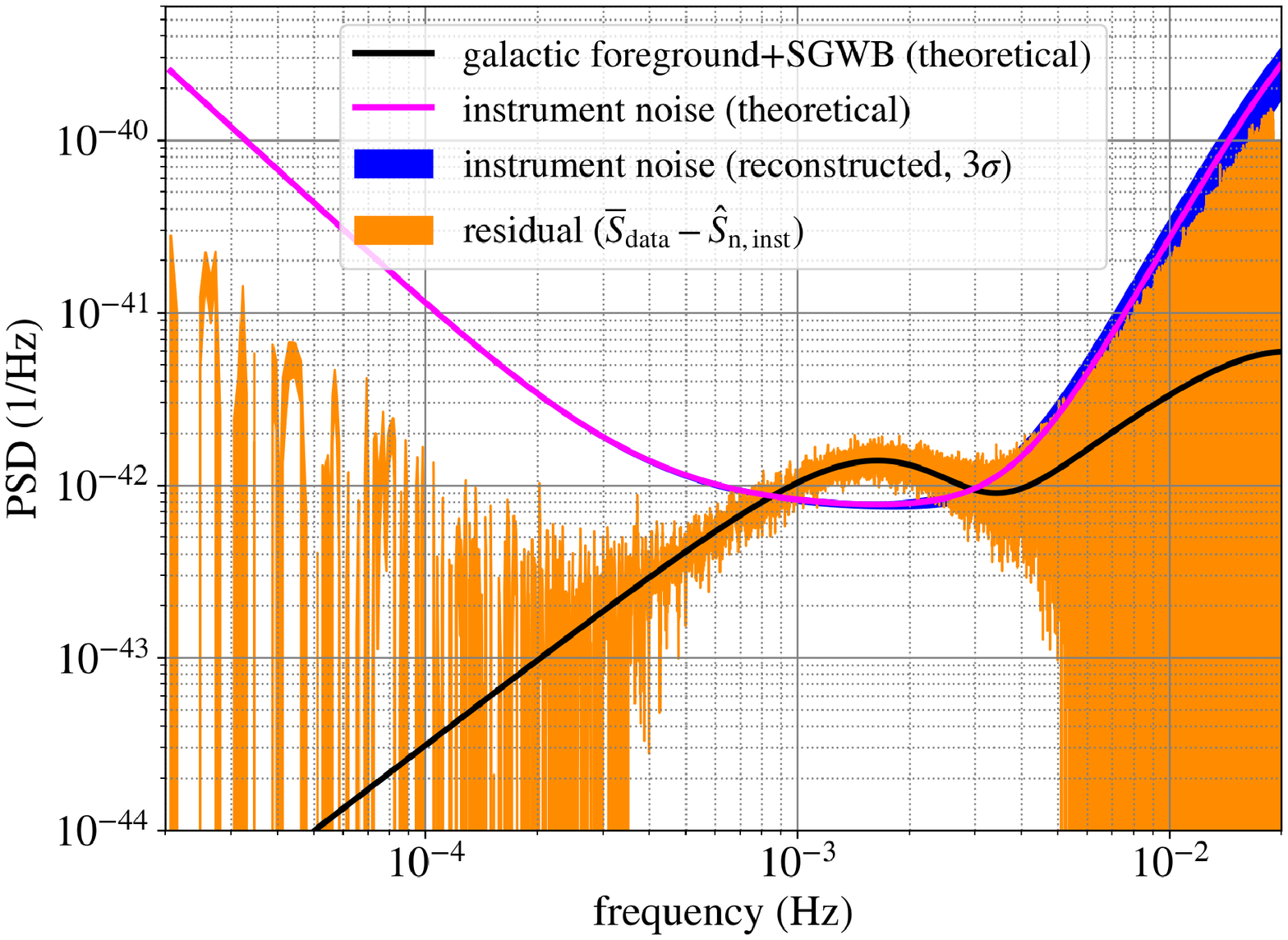}
\caption{The reconstructed spectral shapes of instrument noise and signals in the Michelson-A channel from Michelson+Relay combination for the science case (upper panel) and the noise-only case (lower panel). For curves from the science case, the PSDs of the galactic foreground and SGWB are modeled and their spectral shapes are constructed by substituting the inferred values into Eqs. \eqref{eq:GB_signal} and \eqref{eq:SGWB_signal}. For the noise-only case, only noise PSD is recovered from characterized noise components. By subtracting the reconstructed noise from data, the injected signals emerge from residual at frequencies in which the signal's PSD surpasses the noise. \label{fig:LISA_Sn_reconstructed}
}
\end{figure}

The Michelson-A channel is selected to illustrate the spectrum reconstruction, the restored PSDs of noise and signals in the $3\sigma$ confidence level are shown in the upper panel of Fig. \ref{fig:LISA_Sn_reconstructed}. As the blue area shows, the noise PSD is restored with good precision and matches the theoretical noise curve accurately. Especially for the frequencies higher than $\sim$3 mHz which are dominated by optical path noises, the noise PSD could be stringently constrained by the high parameter resolution from the Michelson+Relay combination. More importantly, since LISA's most sensitive band is around $\sim$10 mHz, the well-characterized noise in this band is valuable to identify any GW signals.
The spectral shape of galactic foreground is also precisely recovered since its amplitude parameter is well constrained. For the SGWB signal, due to the degeneracy between $A_0$ and $\alpha_0$, its spectral shape is recovered more precisely at frequencies of $\sim$4 mHz, and the uncertainties of the PSD increase for frequencies far from the 4 mHz.

The signal reconstruction in the science case presumes that the signals are expectable. A possible scenario that may be encountered is that a GW signal is unforeseen and its spectral shape is unmodelled \cite{Cornish:2019fee}. For the ground-based interferometer network, the unmodelled SGWB could be identified by analyzing multiple data streams from independent detectors \cite{Christensen:1992wi,Allen:1996vm,Allen:1997ad}. Similarly, the signal could also be recognized by using the cross-correlation between two space detectors, such as the LISA-TAIJI network \cite{Omiya:2020fvw,Orlando:2020oko,Wang:2021njt}. As an alternative approach, for single LISA mission, if the PSDs of instrumental noises are sufficiently characterized, the unforeseen signal may be discerned. 

The noise-only case explored noise characterization by combining GW-insensitive data. 
In this case, only PSD of noise is reconstructed from the characterized noise components, and the injected galactic foreground and SGWB are treated as unmodelled signals. The reconstructions for the Michelson-A channel are shown in the lower panel of Fig. \ref{fig:LISA_Sn_reconstructed}. The blue curve shows the reconstructed noise PSD in a $3\sigma$ confidence interval. Compared to the theoretical curve shown by magenta, the noise PSD could be recovered with higher precision at lower frequencies which are dominated by acceleration noises. In the higher frequency band, the PSD of noise is reconstructed with larger uncertainties because the optical path noise is loosely determined. By subtracting the restored noise PSD from data, the spectrum of residual is shown by the yellow area. In the noise-subtracted residual, the spectral shape of the signal could be recognized at frequencies in which signal's PSD surpasses the noise, and it matches the total PSD of injected two signals shown by the black curve. In contrast, the signal can not be discriminated once the PSD of signal is beneath the noise curve. Even so, an upper limit constraint could be obtained in the lower frequency band since the acceleration noise is relatively better estimated. And in the higher frequencies, the upper boundary of the signal would be limited by the uncertainties of the noise PSD. 

Comparing the two plots in Fig. \ref{fig:LISA_Sn_reconstructed}, the spectra reconstructed from the science case are more precise than the noise-only case. The first reason is that the science case employs a larger frequency band of regular TDI channels to achieve a better noise characterization. And the second reason is that parameters of modeled signals are directly inferred from data and the PSDs are reconstructed by substituting the inferred values into Eqs. \eqref{eq:GB_signal} and \eqref{eq:SGWB_signal}. Although the PSDs are reconstructed with less precision from the noise-only combination, it may be still enough to identify a GW signal whose power exceeds the noise level. And this generic approach could mitigate the model dependence for an unexpected GW signal.

\section{Conclusions} \label{sec:conclusions}

In this work, we perform the noise characterizations by combining TDI channels from the first-generation configurations.
Although the T channel, as a null stream, is promising to characterize noises, its loose constraints on noise parameters may not satisfy the requirements. Moreover, the optical path noise overwhelms the acceleration noise in the T channel especially for the equal arm cases, and additional data is needed to characterize the noises thoroughly. The combinations are explored between the first-generation TDI configurations to achieve the synergy for noise determination. 
The TDI channels from Relay configuration could effectively solve the degeneracy between optical path noises, and the combination of Michelson and Relay could efficiently characterize instrumental noises. Furthermore, with the characterized noises, the PSD of noise in a TDI channel could be reconstructed, and a GW-induced spectral shape may be recognized.

To demonstrate the characterization of the GW signals, the galactic foreground and a power-law spectral shape SGWB are simulated. From an optimistic perspective, if the signal is predictable, its parameter(s) could be directly estimated from the TDI data, and its spectral shape could be restored from the inferred values. From a pessimistic or more generic perspective, if the spectrum of an SGWB is unforeseen, then the well-characterized noises will be valuable for signal identification. As we examined in the noise-only case, the unexpected signal could be recognized and retrieved from noise-subtracted residual if its PSD exceeds the noise level. The galactic foreground, as a confusion noise for the LISA, surpasses the instrumental noise around 1 mHz, and it may be separated and removed from data by adopting this approach.
Caveats for this investigation: the SGWB is presumed unlikely to be observed in the very low frequencies for LISA, and the GW-insensitive data in this band is utilized to characterize the noises. If the signal is present in the low-frequency band, alternative algorithms should be developed depending on the significance of the SGWB observation. The noise characterizations are performed by utilizing the pre-known noise spectral shapes, and the noise shapes may differ from the real mission operation. The noise analysis for LISA Pathfinder would improve our understanding of the noise models for the final mission \cite{Castelli:2020zro}.

In this investigation, we presume that all noises are Gaussian and stationary during three-year observation and the stochastic GW signals are isotropic. During the reality operation, non-stationary noise or glitches may happen, and the impact of glitches could also be mitigated by a TDI combination \cite{Robson:2018jly}. For an anisotropic stochastic signal, we also expect its spectrum could also be separated from the well-characterized instrumental noises. 
On the other side, the first-generation TDI configurations, Michelson and Relay, are employed to characterize noises in this work. Their second-generation configurations could be required in realistic observation. Considering the second-generation TDI channel is constructed from two time-shifted first-generation channels, the coefficients of noise components are multiplied by a constant sinusoidal factor \cite{Krolak:2004xp}. The noise characterization achieved from the first-generation configurations may also be anticipated from the second-generation Michelson and Relay combination, and the relevant study will be fulfilled in our future work.

\begin{acknowledgments}

This work was supported by the National Key R\&D Program of China under Grant Nos. 2021YFC2201903 and 2020YFC2201400, and NSFC Nos. 12003059, 11933010, 11873097 and 11922303. This work made use of the High Performance Computing Resource in the Core Facility for Advanced Research Computing at Shanghai Astronomical Observatory.
G.W. thanks Zhen Yan for insightful discussions for the parameter estimations. 
The calculations in this work are performed by using the python packages $\mathsf{numpy}$ \cite{harris2020array} and $\mathsf{scipy}$ \cite{2020SciPy-NMeth}, and the plots are make by utilizing $\mathsf{matplotlib}$ \cite{Hunter:2007ouj}, $\mathsf{Component Library}$ \cite{ComponentLibrary} and $\mathsf{GetDist}$ \cite{Lewis:2019xzd}.

\end{acknowledgments}

\appendix

\section{Response formulation of laser link to GW} \label{sec:appendix_response}

For a source locating at ecliptic longitude $\lambda$ and latitude $\theta$ (in the solar-system barycentric coordinates), the GW propagation vector will be
\begin{equation} \label{eq:source_vec}
 \hat{k}  = -( \cos \lambda \cos \theta, \sin \lambda \cos \theta ,  \sin \theta ).
\end{equation}
The $+$, $\times$ polarization tensors of the GW signal combining source's inclination angle $\iota$ are
\begin{equation} \label{eq:polarizations-response}
\begin{aligned}
{\rm e}_{+} & \equiv \mathcal{O}_1 \cdot
\begin{pmatrix}
1 & 0 & 0 \\
0 & -1 & 0 \\
0 & 0 & 0
\end{pmatrix}
\cdot \mathcal{O}^T_1 \times \frac{1+\cos^2 \iota}{2} ,
\\
{\rm e}_{\times} &  \equiv \mathcal{O}_1 \cdot
\begin{pmatrix}
0 & 1 & 0\\
1 & 0 & 0 \\
0 & 0 & 0
\end{pmatrix}
\cdot \mathcal{O}^T_1 \times i (- \cos \iota ),
\end{aligned}
\end{equation}
with
\begin{widetext}
\begin{equation}
\mathcal{O}_1 =
\begin{pmatrix}
\sin \lambda \cos \psi - \cos \lambda \sin \theta \sin \psi & -\sin \lambda \sin \psi - \cos \lambda \sin \theta \cos \psi & -\cos \lambda \cos \theta  \\
     -\cos \lambda \cos \psi - \sin \lambda \sin \theta \sin \psi & \cos \lambda \sin \psi - \sin \lambda \sin \theta \cos \psi & -\sin \lambda \cos \theta  \\
         \cos \theta \sin \psi & \cos \theta \cos \psi & -\sin \theta
\end{pmatrix},
\end{equation}
where $\psi$ is polarization angle. The response to the GW in laser link from S/C$i$ to $j$ will be
\begin{equation} \label{eq:y_ij}
\begin{aligned}
y^{h}_{ij} (f) =&  \frac{ \sum_\mathrm{p} \hat{n}_{ij} \cdot {\mathrm{ e_p}} \cdot \hat{n}_{ij} }{2 (1 - \hat{n}_{ij} \cdot \hat{k} ) }
 \times \left[  \exp( 2 \pi i f (L_{ij} + \hat{k} \cdot p_i ) ) -  \exp( 2 \pi i f  \hat{k} \cdot p_j )  \right] ,
\end{aligned}
\end{equation}
\end{widetext}
where $\hat{n}_{ij}$ is the unit vector from S/C$i$ to $j$, $L_{ij}$ is the arm length from S/C$i$ to $j$, $p_i$ is the position of the S/C$i$ in the solar-system barycentric ecliptic coordinates. 

\section{Independence of Relay from Sagnac for optical metrology noise characterization} \label{sec:Sagnac_Relay}

The TDI observables are supposed to be linear combinations of a set of generators ($\alpha$, $\beta$, $\gamma$, and $\zeta$)  \cite{Dhurandhar:2002zcl,Tinto:2020fcc}. However, the PSD of a TDI data stream or CSD of two channels could not be linearly composed by the spectral densities of four generators. The TDI data streams beyond generators could have different performances on noise characterization. The Relay configuration is selected to explicate the changes for optical metrology noise characterization. To simplify the expressions, the acceleration noise components are ignored and the interferometric arms are assumed to be equal. Then the Sagnac-$\alpha$ in Eq. \eqref{eq:alpha_expression} could be expressed as,
\begin{equation}
\alpha =  (n^\mathrm{op}_{13} + \mathcal{D} n^\mathrm{op}_{32} + \mathcal{D}^2 n^\mathrm{op}_{21} ) 
 - ( n^\mathrm{op}_{12} +  \mathcal{D} n^\mathrm{op}_{23} + \mathcal{D}^2 n^\mathrm{op}_{31} ).
\end{equation} 
By assuming the different optical metrology noises are independent, and the PSD of a noise component is $S_{\mathrm{op}ij} = \langle | \tilde{n}^\mathrm{op}_{ij} |^2 \rangle $. The PSDs and CSDs between three generators ($\alpha$, $\beta$, and $\gamma$) will be
\begin{widetext}
\begin{center}
$
\begin{bmatrix}
S_{\alpha \alpha} \\
S_{\beta \beta} \\
S_{\gamma \gamma} \\
S_{\alpha \beta} \\
S_{\alpha \gamma} \\
S_{\beta \gamma}
\end{bmatrix}
=
\begin{bmatrix}
1 & 1 & 1 & 1 & 1 & 1\\
1 & 1 & 1 & 1 & 1 & 1\\
1 & 1 & 1 & 1 & 1 & 1 \\
\cos 2 x & \cos x & \cos 2 x & \cos x & \cos x & \cos x \\
\cos x & \cos 2 x & \cos x & \cos x & \cos 2 x & \cos x \\
\cos x & \cos x & \cos x & \cos 2 x & \cos x & \cos 2 x 
\end{bmatrix}
\begin{bmatrix}
S_\mathrm{op12} \\
S_\mathrm{op13} \\
S_\mathrm{op21} \\
S_\mathrm{op23} \\
S_\mathrm{op31} \\
S_\mathrm{op32}
\end{bmatrix}
=
M^\mathrm{op}_\mathrm{Sagnac}
\begin{bmatrix}
S_\mathrm{op12} \\
S_\mathrm{op13} \\
S_\mathrm{op21} \\
S_\mathrm{op23} \\
S_\mathrm{op31} \\
S_\mathrm{op32}
\end{bmatrix}.
$
\end{center}
\end{widetext}
The coefficient matrix, $M^\mathrm{op}_\mathrm{Sagnac}$, is singular which means PSDs and CSDs from three Sagnac observables could not separately resolve the PSDs of six noise components. On the other side, the Relay data streams are linear combinations of the Sagnac observables, for instance, Relay-U \cite{Tinto:2020fcc},
\begin{equation}
\tilde{ \mathrm{U} } = \mathcal{D} \tilde{ \gamma } - \tilde{ \beta } = e^{ i x} \tilde{ \gamma } - \tilde{ \beta }.
\end{equation}
Its PSD would be
\begin{equation}
S_\mathrm{UU} = \langle \tilde{U} \tilde{U}^\ast \rangle = S_{\gamma \gamma} + S_{\beta \beta } - 2 \Re \langle e^{ i x} \tilde{ \gamma } \tilde{ \beta }^\ast \rangle.
\end{equation}
We can realize the last term on the right side, $\Re \langle e^{ i x} \tilde{ \gamma } \tilde{ \beta }^\ast \rangle$, may not be a linear term of $S_{\gamma \beta} $, and $S_\mathrm{UU}$ could not be linearly composed by the spectral densities of Sagnac channels. Specifically, the PSDs and CSDs of Relay observables will be,
\begin{widetext}
\begin{center}
$
\begin{bmatrix}
S_\mathrm{UU} \\
S_\mathrm{VV} \\
S_\mathrm{WW} \\
S_\mathrm{UV} \\
S_\mathrm{UW} \\
S_\mathrm{VW}
\end{bmatrix}
=
\begin{bmatrix}
  0 & 4 \sin^2 x  &  4 \sin^2 x  &  4 \sin^2 \frac{3x}{2} & 0 &  4 \sin^2 \frac{x}{2}  \\
  0 & 4 \sin^2 \frac{x}{2} & 4 \sin^2 x & 0 & 4 \sin^2 \frac{3x}{2} & 4 \sin^2 x  \\
  4 \sin^2 \frac{3x}{2} & 4 \sin^2 x  & 4 \sin^2 \frac{x}{2} & 0 & 0 & 4 \sin^2 x  \\
  0 & -2 \sin^2 x  &  4 \cos x \sin^2 x & 0 & 0 & -2 \sin^2 x \\
  0 & 4 \cos x \sin^2 x & -2 \sin^2 x  & 0 & 0 & -2 \sin^2 x \\
  0 & -2 \sin^2 x  & -2 \sin^2 x &  0 & 0 &  4 \cos x \sin^2 x \\
\end{bmatrix}
\begin{bmatrix}
S_\mathrm{op12} \\
S_\mathrm{op13} \\
S_\mathrm{op21} \\
S_\mathrm{op23} \\
S_\mathrm{op31} \\
S_\mathrm{op32}
\end{bmatrix}
=
M^\mathrm{op}_\mathrm{Relay}
\begin{bmatrix}
S_\mathrm{op12} \\
S_\mathrm{op13} \\
S_\mathrm{op21} \\
S_\mathrm{op23} \\
S_\mathrm{op31} \\
S_\mathrm{op32}
\end{bmatrix},
$
\end{center}
\end{widetext}
where the coefficient matrix, $M^\mathrm{op}_\mathrm{Relay}$, is invertible when $x \neq n \pi, \ (\mathrm{for} \ n=1, 2, 3...)$. Therefore, (except for particular frequencies), the Relay data streams could break the degeneracies between optical metrology noises and characterize these components individually. As we can conclude, although the Relay data streams could be yielded from Sagnac, the Relay would have better capacities for optical metrology noise characterization.
Moreover, when the optimal channels are obtained from regular channels by applying Eq. \eqref{eq:optimalTDI}, their spectral densities of optimal channels are linear transformation from spectral densities of regular channels \cite{Prince:2002hp}, and the singular/nonsingular property of the coefficient matrix will not be changed. The noise characterization from three optimal channels should be consistent with the results from the corresponding three regular data streams.

\nocite{*}
\bibliography{apsref}

\end{document}